 \documentclass[12pt]{article}

\usepackage{amsfonts}
\usepackage{amssymb}
\usepackage{amsmath}
\usepackage{latexsym}
\usepackage{verbatim}
\usepackage[english]{babel}

\allowdisplaybreaks[1]

\newcommand{\overskrift}[1]{\vspace{3.0mm}\noindent\textbf{#1}\vspace{1.5mm}}
\newcommand{\storoverskrift}[1]{\vspace{3.0mm}\noindent{\Large\textbf{#1}}\vspace{1.5mm}}
\newcommand{\vp}{\varphi}
\newcommand{\dphi}{\delta^{\textrm{\tiny (1)}} \varphi}
\newcommand{\ddphi}{\delta^{\textrm{\tiny (2)}} \varphi}
\newcommand{\dsigma}{\delta^{\textrm{\tiny (1)}} \sigma}
\newcommand{\ddsigma}{\delta^{\textrm{\tiny (2)}} \sigma}
\newcommand{\dphid}{\delta^{\textrm{\tiny (1)}} \dot{\varphi}}

\newcommand{\dsigmad}{\delta^{\textrm{\tiny (1)}} \dot{\sigma}}

\newcommand{\del}{\delta^{\textrm{\tiny (1)}}}
\newcommand{\ddel}{\delta^{\textrm{\tiny (2)}}}
\newcommand{\p}{\partial}
\newcommand{\Rcal}{{\cal R}}
\newcommand{\Pcal}{{\cal P}}
\newcommand{\Hcal}{{\cal H}}
\newcommand{\Ocal}{{\cal O}}
\newcommand{\Ical}{{\cal I}}
\newcommand{\Kcal}{{\cal K}}
\newcommand{\ixo}{\textrm{\tiny (1)}}
\newcommand{\ixt}{\textrm{\tiny (2)}}
\newcommand{\ix}[1]{\textrm{\tiny (#1)}}
\newcommand{\nn}{\nonumber}
\newcommand{\GeV}{\rm GeV}
\newcommand{\largedot}{\mbox{\boldmath $\cdot$}}
\newcommand{\largeddot}{\mbox{\boldmath $\cdot \cdot$}}

\begin{document}

 \begin{titlepage}
 \begin{flushleft}
        \hfill                      {\tt hep-ph/0405103}\\ \hfill
        HIP-2004-22/TH \\ \hfill
 \end{flushleft}
 \vspace*{3mm}
 \begin{center}
 {\Large { \bf Non-Gaussian perturbations in hybrid inflation \\}}
 \vspace*{12mm} { \large Kari
 Enqvist\footnote{E-mail: kari.enqvist@helsinki.fi} and
 Antti V\"aihk\"onen\footnote{E-mail: antti.vaihkonen@helsinki.fi}\\}

\vspace{5mm}

 {
 Helsinki Institute of Physics and
 Department of Physical Sciences \\ P.O. Box 64, FIN-00014
 University of Helsinki, Finland}

\vspace*{10mm}
\end{center}


\begin{abstract} \noindent
  We consider second order inflationary perturbations in the case of two
  scalar fields, $\sigma$ and the inflaton $\vp$. We derive an expression for
  the non-Gaussianity of perturbations and apply the results to hybrid
  inflation. We isolate the contributions due to $\sigma$ and evaluate the
  resulting terms to show that $\sigma$-induced non-Gaussianities dominate
  over inflaton-induced non-Gaussianities when $m^2_{\sigma} \gtrsim \eta
  H^2$, where $\eta$ is the slow-roll parameter.  This may provide a useful
  test of hybrid inflation in forthcoming CMB experiments.

\end{abstract}

\end{titlepage}

\baselineskip16pt


\section{Introduction}

With the advent of WMAP \cite{wmap}, measurements of the temperature and
polarization fluctuations of the cosmic microwave background (CMB) now begin
to reach the level of precision needed for testing various models of inflation
and hence, by implication, particle physics at very short length scales. The
spectral index of the two-point correlator of scalar perturbations is a
well-known measure of the slow-roll parameters, which are directly related to
the form of the inflaton potential. Tensor perturbations, the spectrum of
which could be extracted from future CMB polarization and temperature
fluctuation data, would yield independent constraints on the potential
\cite{knox03,wands02}. In case of multi-field inflaton models \cite{lyth99a},
testing is naturally more involved. For instance, the constraints imposed by
the tensor modes are less severe \cite{wands02}. Moreover, in addition to the
purely adiabatic perturbations of the single-field inflation, there could be a
small component of entropy (isocurvature) perturbation (for a general
treatment of the first order entropy perturbations, see \cite{gordon01}).
There could also be a correlation between the adiabatic and entropy
perturbations which is a source for additional freedom in fitting the CMB data
\cite{valiviita03}.

In addition to the spectral features of the CMB, the statistical properties of
the temperature fluctuations also provide potentially important information
about the origin of the primordial perturbations. This is an aspect that has
received much less attention as compared to the spectral considerations -- and
up to now, perhaps for a good reason. From observations we know that the
spatial distribution of the temperature fluctuations on the microwave sky
appear to be random so that their statistics is mostly Gaussian
\cite{wmap:gaussianity}. However, in many inflationary models there are
several sources of non-Gaussianities; the self-interaction of the inflaton is
one obvious example \cite{maldacena03}. In models with more than one scalar
field the emergence of non-Gaussianity at some level cannot usually be
avoided. These non-Gaussianities may be small, but they could nevertheless
yield important constraints on models of inflation.  Indeed, observations of
CMB non-Gaussianity are only now starting to be sensitive enough to be able to
distinguish between different inflationary scenarios \cite{komatsu01} (see
also \cite{komatsu02} and references therein).

Non-Gaussianity in standard single-field inflation has been considered e.g.\ 
within the framework of stochastic inflation \cite{salopek90,salopek91}. The
three-point correlation function was discussed in \cite{gangui94}. For
Gaussian statistics, all the $N$-point correlators are related to the
two-point correlator with the odd correlators vanishing. Hence the three-point
correlator of the scalar perturbations is the lowest order measure of
non-Gaussianity of the perturbations.  In \cite{maldacena03} the induced cubic
self-interaction terms of the inflaton were computed together with an estimate
for the resulting non-Gaussianity, which was found to be small. Non-vacuum
initial states and non-Gaussianity have been addressed in
\cite{lesgourgues96,martin99,gangui02}.  Nonstandard inflation scenarios have
also been studied in the context of non-Gaussianity, such as models with
higher order derivative operators \cite{creminelli03}, or models with a
varying inflaton decay rate \cite{zaldarriaga03}. However, many of these
treatments (with the exception of \cite{maldacena03}) have been lacking in
that non-Gaussianity has been introduced by hand, often relating it to the
magnitude of the first-order perturbation. In general a consistent treatment
of the perturbation theory up to second order is required to find the Fourier
transform of the three-point correlator, called the bispectrum.

A full computation of the bispectrum of single-field inflationary models has
recently been performed in second-order perturbation theory by Acquaviva et
al.\ \cite{acquaviva03}. They considered second-order perturbations in both
the metric and the energy-momentum tensor and, using the Einstein equation,
solved the evolution of the gauge-invariant comoving curvature perturbation,
computed up to the first order in the slow-roll parameters. This enabled them
to find out the gauge-invariant gravitational potential bispectrum.

Despite the occasional incompleteness of their treatment, most single-field
studies suggest that the amount of primordial non-Gaussianity is unobservable
at present (see however \cite{zaldarriaga03}) and possibly unobservable even
by the Planck satellite \cite{komatsu01}. However, many models of inflation
are not based on the simple single-field approach. Indeed, one of the most
popular realization of the inflationary paradigm is hybrid inflation
\cite{linde91} which relies on two scalar fields, the slowly rolling inflaton
$\vp$, and an additional field $\sigma$, sometimes called the ``waterfall
field''. During inflation $\sigma$ stays roughly constant but eventually
triggers the end of inflation by rapidly cascading down the potential to its
minimum (see e.g.\ \cite{linde94,lyth99a}).

Non-Gaussianity in multi-field inflation has been studied by Bernardeau and
Uzan \cite{bernardeau02,bernardeau03} (even before them Yi and Vishniac
\cite{yi93} concluded that non-Gaussian effects can be important in
multi-field models but their model lacked compelling physical motivation). In
their case inherently non-Gaussian isocurvature perturbations are transferred
to the adiabatic modes during inflation. The necessary ingredients for this
kind of mechanism are a self-coupling for a transverse scalar field for
generating the initial non-Gaussianity, and a coupling in the potential
leading to a curved trajectory in the field space (and trough that to a mixing
between isocurvature and adiabatic modes). This approach is very specific and
is not applicable to a generic multi-field inflation. Indeed, as in the case
of single-field inflation, the proper treatment of non-Gaussianity requires
expanding metric and scalar field perturbations up to second order. Second
order perturbation theory \cite{bruni97,matarrese98,noh03} has been employed
both in single-field models \cite{maldacena03,acquaviva03} and multi-field
models \cite{bartolo03a}, where the emphasis was on the connection between
primordial perturbations and CMB temperature fluctuation at large scales.  In
\cite{rigopoulos02} it was applied to a study of isocurvature and adiabatic
perturbations in multi-field models (see also \cite{bartolo01}), where it was
found that the isocurvature perturbation sources the gravitational potential
on long wavelengths in the second order even without any bend in the
trajectory in the field space. We reach the same conclusion in our study.
Second order perturbation theory has also been applied to the curvaton
scenario \cite{enqvist01, lyth01, moroi01}, where the curvature perturbations
are generated after the inflation; for a discussion on non-Gaussianity in
curvaton models, see \cite{lyth02,bartolo03a,bartolo03c}.  The enhancement of
primordial perturbations after inflation has been studied in
\cite{bartolo03b}, where it was found that even tiny non-Gaussianities
produced during single-field inflation could be significantly enhanced by
gravitational dynamics after inflation.

This paper is organized as follows. In Section 2 we expand both
the metric and energy-momentum tensor perturbations of two scalar
fields, denoted as $\vp$ (the inflaton) and $\sigma$, up to second
order, following the single-field treatment presented in
\cite{acquaviva03}. We then write down the perturbed Einstein
equations up to second order and construct a master equation for
the second order metric perturbation $\phi^{\ixt}$. In Section 3
we apply the master equation to hybrid inflation. Section 4 is
devoted to constructing second order comoving curvature
perturbation first for a general two-field case and then for
hybrid inflation. The contribution due to the second scalar field
$\sigma$ is isolated. In Section 5 we study the curvature
perturbation induced by the second field in more detail. In
particular the amount of non-Gaussianity due to $\sigma$ is
compared to non-Gaussianity due to inflaton. Finally in Section 6
we discuss our results and draw conclusions.

\section{Second order inflationary perturbations of two\\ scalar fields}

\subsection{Metric perturbations}

Let us start by assuming a spatially flat Robertson-Walker metric in conformal
time $ds^2=g_{\mu\nu}dx^\mu dx^\nu = a^2(\tau)(-d\tau^2+d{\boldsymbol x}^2)$.
The components of a perturbed metric up to an arbitrary order can be written
as \cite{bruni97,matarrese98}
\begin{eqnarray}
  g_{00} &=& -a(\tau)^2 \left( 1 + 2 \sum^{\infty}_{r=1} \frac{1}{r!}
  \phi^{\ix r} \right), \\
  g_{0i} &=& a(\tau)^2 \sum^{\infty}_{r=1} \frac{1}{r!} \hat{\omega}_i^{\ix
  r}, \\
  g_{ij} &=& a(\tau)^2 \left[ \left( 1 - 2 \sum^{\infty}_{r=1} \frac{1}{r!}
  \psi_i^{\ix r} \right) \delta_{ij} + \sum^{\infty}_{r=1} \frac{1}{r!}
  \hat{\chi}^{\ix r}_{ij} \right],
\end{eqnarray}
where the functions $\phi^{\ix r}$, $\hat{\omega}^{\ix r}_i$, $\psi^{\ix r}$,
and $\hat{\chi}^{\ix r}_{ij}$ represent the $r$th order perturbations of the
metric.

Let us perform the standard splitting into the scalar, vector, and tensor
parts by writing \cite{bruni97,matarrese98}
\begin{eqnarray}
  \hat{\omega}_i^{\ix r} &=& \p_i \omega^{\ix r} + \omega_i^{\ix r}, \\
  \hat{\chi}^{\ix r}_{ij} &=& D_{ij} \chi^{\ix r} + \p_i \chi_j^{\ix r} + \p_j
  \chi_i^{\ix r} + \chi_{ij}^{\ix r},
\end{eqnarray}
where $\p^i \omega_i^{\ix r} = \p^i \chi_i^{\ix r} = 0$, $\p^i \chi_{ij}^{\ix
  r} = 0$, $\chi^{i \ix r}_{\phantom i i}=0$, and $D_{ij} = \p_i \p_j -
\frac{1}{3} \delta_{ij} \p^k \p_k$. When considering only scalar fields up to
the second order we can follow Acquaviva et al.\ \cite{acquaviva03} and
neglect $\omega_i^{\ixo}$, $\chi_i^{\ixo}$, and $\chi_{ij}^{\ixo}$, thus
obtaining the metric
\begin{eqnarray}
  g_{00} &=& -a(\tau)^2 \left( 1 + 2 \phi^{\ixo} + \phi^{\ixt} \right), \\
  g_{0i} &=& a(\tau)^2 \left( \p_i \omega^{\ixo} + \frac{1}{2} \p_i
  \omega^{\ixt} + \frac{1}{2} \omega_i^{\ixt} \right), \\
  g_{ij} &=& a(\tau)^2 \left[ \left( 1 - 2 \psi^{\ixo} - \psi^{\ixt} \right)
  \delta_{ij} + D_{ij} \left( \chi^{\ixo} + \frac{1}{2} \chi^{\ixt} \right)
  \right. \nn \\
  & & \left. + \frac{1}{2} \left( \p_i \chi_j^{\ixt} + \p_j \chi_i^{\ixt} +
  \chi_{ij}^{\ixt} \right) \right].
\end{eqnarray}
We will adopt the generalized longitudinal gauge \cite{bruni97,matarrese98},
and set $\omega^{\ixo} = \omega^{\ixt} = \omega_i^{\ixt} = 0$ and $\chi^{\ixo}
= \chi^{\ixt} = 0$.  This renders the metric into the form
\begin{eqnarray}
  g_{00} &=& -a(\tau)^2 \left( 1 + 2 \phi^{\ixo} + \phi^{\ixt} \right),
  \label{eq:metric00}\\
  g_{0i} &=& 0~, \label{eq:metric0i}\\
  g_{ij} &=& a(\tau)^2 \left[ \left( 1 - 2 \psi^{\ixo} - \psi^{\ixt} \right)
  \delta_{ij} + \frac{1}{2} \left( \p_i \chi_j^{\ixt} + \p_j \chi_i^{\ixt} +
  \chi_{ij}^{\ixt} \right) \right]~, \label{eq:metricij}
\end{eqnarray}
which is the basis of our subsequent analysis. The Einstein tensor resulting
from the metric Eqs.\ (\ref{eq:metric00})-(\ref{eq:metricij}) is presented in
Appendix \ref{app:einstein}.

\subsection{Energy-momentum perturbations of two scalar fields}

Let us now consider the energy-momentum tensor $T_{\mu \nu}$ for
generic two scalar fields, $\vp$ and $\sigma$, minimally coupled
to gravity and given by \cite{liddle00}
\begin{equation}
  T_{\mu \nu} = \p_{\mu} \vp \,\p_{\nu} \vp + \p_{\mu} \sigma \,\p_{\nu}
  \sigma - g_{\mu \nu} \left( \frac{1}{2} \,g^{\alpha \beta} \,\p_{\alpha} \vp
  \,\p_{\beta} \vp + \frac{1}{2} \,g^{\alpha \beta} \,\p_{\alpha} \sigma
  \,\p_{\beta} \sigma + V(\vp,\sigma) \right), \label{eq:energymomentumtensor}
\end{equation}
where $V(\vp,\sigma)$ is the potential for the scalar fields. In our notation,
$\vp$ is the inflaton. The fields can be
expanded up to the second order in perturbations, where the zeroth order is
the homogeneous part denoted by the subscript $0$:
\begin{eqnarray}
  \vp(\tau, \boldsymbol x) &=& \vp_0(\tau) + \dphi(\tau, \boldsymbol x) +
  \frac{1}{2} \ddphi(\tau, \boldsymbol x), \label{eq:phi-expansion}\\
  \sigma(\tau, \boldsymbol x) &=& \sigma_0(\tau) + \dsigma(\tau, \boldsymbol
  x) + \frac{1}{2} \ddsigma(\tau, \boldsymbol x). \label{eq:sigma-expansion}
\end{eqnarray}
Similarly, the energy-momentum tensor can be expanded up to second order as
\begin{equation} \label{eq:em-tensor-expansion}
  T^{\mu}_{\phantom m \nu} = T^{\mu \ix 0}_{\phantom m \nu} + \del
  T^{\mu}_{\phantom m \nu} + \frac{1}{2} \ddel T^{\mu}_{\phantom m \nu},
\end{equation}
where $ T^{\mu \ix 0}_{\phantom m \nu}$ denotes the background value. The
components of the energy-momentum tensor (\ref{eq:energymomentumtensor}) up to
the second order are presented in Appendix \ref{app:energymomentumtensor}.

\subsection{Einstein equations}

Using the results of the two previous subsections we may expand the Einstein
tensor up to the second order as\footnote{Note that Acquaviva et al.
  \cite{acquaviva03} expand $G_{\mu \nu} = G_{\mu \nu}^{\ix 0} + \del G_{\mu
    \nu} + \ddel G_{\mu \nu}$ without the factor one half in the second order
  term, although they have it in the expansion of the energy-momentum tensor.}
\begin{equation} \label{eq:einstein-exp}
  G_{\mu \nu} = G_{\mu \nu}^{\ix 0} + \del G_{\mu \nu} + \frac{1}{2} \ddel
  G_{\mu \nu}.
\end{equation}
We list the components of the Einstein tensor for the metric
(\ref{eq:metric00}) - (\ref{eq:metricij}) up to the second order in Appendix
\ref{app:einstein}.

The Einstein equations can now be written in the component form. The
background equations\footnote{Note that $G^{0 \ix 0}_{\phantom 0 i} = G^{i \ix
    0}_{\phantom 0 0} = T^{0 \ix 0}_{\phantom 0 i} = T^{i \ix 0}_{\phantom 0
    0} = 0$.}  $G^{0 \ix 0}_{\phantom 0 0} = \kappa^2 T^{0 \ix 0}_{\phantom 0
  0}$ and $G^{i \ix 0}_{\phantom i j} = \kappa^2 T^{i \ix 0}_{\phantom i j}$
are found to be respectively given by
\begin{equation} \label{eq:einstein-0-00}
  3 \Hcal^2 = \frac{\kappa^2}{2} \left( {\vp'_0}^2 + {\sigma'_0}^2 \right) +
  \kappa^2 a^2 V_0
\end{equation}
and
\begin{equation} \label{eq:einstein-0-ij}
  \Hcal^2 + 2 \Hcal' = -\frac{\kappa^2}{2} \left( {\vp'_0}^2 + {\sigma'_0}^2
  \right) + \kappa^2 a^2 V_0~.
\end{equation}
Here $\Hcal \equiv a'/a$ and $\kappa^2 \equiv 8 \pi G_N \equiv
M_P^{-2}$, where $G_N$ is Newton's constant and $M_P$ is the
reduced Planck mass.

The first order perturbed equations $\del G^0_{\phantom 0 0} = \kappa^2 \del
T^0_{\phantom 0 0}$, $\del G^0_{\phantom 0 i} = \kappa^2 \del T^0_{\phantom 0
  i}$ and $\del G^i_{\phantom i j} = \kappa^2 \del T^i_{\phantom i j}$ take
respectively the forms
\begin{eqnarray} \label{eq:einstein-1storder}
    &&6\Hcal^2 \phi^{\ixo} + 6 \Hcal {\psi^{\ixo}}' - 2 \p_i \p^i \psi^{\ixo}
    \nn \\
  && ~~=\kappa^2 \left[ -\vp'_0 \,\dphi' - \sigma'_0 \,\dsigma' + w^2
    \phi^{\ixo} - a^2 \left( \frac{\p V}{\p \vp}
      \dphi + \frac{\p V}{\p \sigma} \dsigma \right) \right]~, \nn \\
\null\\ \nn
  &&2 \Hcal \p_i \phi^{\ixo} + 2 \p_i {\psi^{\ixo}}' = \kappa^2 \left( \vp'_0
  \,\p_i \dphi + \sigma'_0 \,\p_i \dsigma \right)~,\\ \nn
\null\\ \nn
  &&\left( 2 \Hcal {\phi^{\ixo}}' + 4 \frac{a''}{a} \phi^{\ixo} - 2 \Hcal^2
    \phi^{\ixo} + \p_k \p^k \phi^{\ixo} + 4 \Hcal {\psi^{\ixo}}' + 2
    {\psi^{\ixo}}'' - \p_k \p^k \psi^{\ixo} \right) \delta^i_{\phantom i j} \\
    \nn
  &&\mbox{} + \p^i \p_j \left( \psi^{\ixo} - \phi^{\ixo} \right) = \kappa^2
    \left[ \vp'_0 \dphi' + \sigma'_0 \dsigma' - w^2 \phi^{\ixo}
    - a^2 \left(\frac{\p V}{\p \vp} \dphi + \frac{\p
    V}{\p \sigma} \dsigma \right) \right] \delta^i_{\phantom i j}~,
\end{eqnarray}
where we have defined $w^2\equiv {\vp'_0}^2+ {\sigma'_0}^2$.

In the second order perturbed equations we use the well known
\cite{mukhanov92} first order result for the scalar fields and set
$\psi^{\ixo}=\phi^{\ixo}$. The components $\ddel G^0_{\phantom 0 0} = \kappa^2
\ddel T^0_{\phantom 0 0}$, $\ddel G^i_{\phantom i 0} = \kappa^2 \ddel
T^i_{\phantom i 0}$ and $\ddel G^i_{\phantom i j} = \kappa^2 \ddel
T^i_{\phantom i j}$ read respectively as
\begin{eqnarray} \label{eq:einstein-2ndorder}
  && \Hcal^2 \phi^{\ixt} + \frac{a''}{a} \phi^{\ixt} + 3\, \Hcal
  {\psi^{\ixt}}' - \p_i \p^i \psi^{\ixt} - 12\, \Hcal^2 \left( \phi^{\ixo}
  \right)^2 - 3\, \p_i \phi^{\ixo} \p^i \phi^{\ixo} \nn\\
  && \mbox{} - 8\, \phi^{\ixo} \p_i \p^i \phi^{\ixo} - 3 \left( {\phi^{\ixo}}'
  \right)^2 = \kappa^2 \left\{ -\frac{1}{2} \left( \vp'_0 \ddphi' + \sigma'_0
  \ddsigma' \right) - \frac{1}{2} \left( \left( \dphi' \right)^2 + \left(
  \dsigma' \right)^2 \right) \right. \nn\\
  && \mbox{} + 2 \left( \vp'_0 \dphi' + \sigma'_0 \dsigma' \right) \phi^{\ixo}
  - 2 w^2 \left( \phi^{\ixo} \right)^2 -
  \frac{1}{2} \left( \p_i \dphi \,\p^i \dphi + \p_i \dsigma \,\p^i \dsigma
  \right) \nn\\
  &&  \left. \mbox{} - \frac{a^2}{2} \left[ \frac{\p V}{\p \vp} \ddphi +
  \frac{\p V}{\p \sigma} \ddsigma + \frac{\p^2 V}{\p \vp^2} \left( \dphi
  \right)^2 + \frac{\p^2 V}{\p \sigma^2} \left( \dsigma \right)^2 + 2
  \frac{\p^2 V}{\p \vp \,\p \sigma} \dphi \dsigma \right] \right\}~, \nn\\
\null \nn\\
  && \Hcal \p^i \phi^{\ixt} + \p^i {\psi^{\ixt}}' + \frac{1}{4} \p_k \p^k
  {\chi^{i \ixt}}' + 2 {\phi^{\ixo}}' \p^i \phi^{\ixo} + 8 \phi^{\ixo} \p^i
  {\phi^{\ixo}}' \nn\\
  &&= \kappa^2 \left[ \frac{1}{2} \left( \vp'_0 \,\p^i \ddphi + \sigma'_0
  \,\p^i \ddsigma \right) + \dphi' \,\p^i \dphi + \dsigma' \,\p^i \dsigma
  \phantom{\frac{1}{2}} \right.\nn\\
  && \left. \phantom{\frac{1}{2}} + 2 \left( \vp'_0 \,\p^i \dphi + \sigma'_0
  \,\p^i \dsigma \right) \phi^{\ixo} \right]~, \\
\null \nn\\
  &&\left[ \frac{1}{2} \p_k \p^k \phi^{\ixt} + \Hcal {\phi^{\ixt}}' +
    \frac{a''}{a} \phi^{\ixt} + \Hcal^2 \phi^{\ixt} - \frac{1}{2} \p_k \p^k
    \psi^{\ixt} + {\psi^{\ixt}}'' \right. \nn\\
  && \mbox{} + 2 \Hcal {\psi^{\ixt}}' + 4 \Hcal^2 \left( \phi^{\ixo} \right)^2
    - 8 \frac{a''}{a} \left( \phi^{\ixo} \right)^2 - 8 \Hcal \phi^{\ixo}
  {\phi^{\ixo}}' - 3 \p_k \phi^{\ixo} \,\p^k \phi^{\ixo} \nn\\
  &&  \left. - \phantom{\frac{1}{2}} 4 \phi^{\ixo} \,\p_k \p^k \phi^{\ixo} -
    \left( {\phi^{\ixo}}' \right)^2 \right] \delta^i_{\phantom i j} -
  \frac{1}{2} \p^i \p_j \phi^{\ixt} + \frac{1}{2} \p^i \p_j \psi^{\ixt} \nn\\
  && \mbox{} + \frac{1}{2} \Hcal \left( \p^i {\chi_j^{\ixt}}' + \p_j {\chi^{i
    \ixt}}' + {\chi^{i \ixt}_{\phantom i j}}' \right) + \frac{1}{4} \left(
    \p^i {\chi_j^{\ixt}}'' + \p_j {\chi^{i \ixt}}'' + {\chi^{i \ixt}_{\phantom
    i j}}'' \right) \nn\\
  && \mbox{} - \frac{1}{4} \p_k \p^k \chi^{i \ixt}_{\phantom i j} + 2\, \p^i
    \phi^{\ixo} \,\p_j \phi^{\ixo} + 4\, \phi^{\ixo} \,\p^i \p_j \phi^{\ixo}
    \nn\\
  && \mbox{} = \kappa^2 \left\{ \left[ \frac{1}{2} \left( \vp'_0 \,\ddphi' +
    \sigma'_0 \,\ddsigma' \right) + \frac{1}{2} \left( \left( \dphi' \right)^2
    + \left( \dsigma' \right)^2 \right) \right. \right. \nn\\
  && \mbox{} - 2 \left( \vp'_0 \dphi' + \sigma'_0 \dsigma' \right) \phi^{\ixo}
    + 2 w^2 \left( \phi^{\ixo} \right)^2
    \nn\\
  && \mbox{} - \frac{1}{2} \left( \p_k \dphi \,\p^k \dphi + \p_k \dsigma
    \,\p^k \dsigma \right) \nn\\
  &&  \left. \mbox{} - \frac{a^2}{2} \left( \frac{\p V}{\p \vp} \ddphi +
    \frac{\p V}{\p \sigma} \ddsigma + \frac{\p^2 V}{\p \vp^2} \left( \dphi
    \right)^2 + \frac{\p^2 V}{\p \sigma^2} \left( \dsigma \right)^2 + 2
    \frac{\p^2 V}{\p \vp \,\p \sigma} \dphi \dsigma \right) \right]
    \delta^i_{\phantom i j} \nn\\
  && \left. \mbox{} + \phantom{\frac{1}{2}} \p^i \dphi \,\p_j \dphi + \p^i
    \dsigma \,\p_j \dsigma \right\}~. \nn
\end{eqnarray}
In writing the $00$ and $ij$ components we have made use of the
equality $\kappa^2 w^2 / 2 = \Hcal^2 - \Hcal'$ of the background
equations (\ref{eq:einstein-0-00}) and (\ref{eq:einstein-0-ij}),
together with the relation $a'' / a = \Hcal^2 + \Hcal'$ which
follows directly from the definition of $\Hcal$.

\subsection{The master equation}

Here we follow the procedure outlined in \cite{acquaviva03} for the single
field case.  Our purpose is to derive master equations for the second order
metric perturbation $\phi^{\ixt}$.  To this end, we take the divergence of the
$i0$ component of the second order Einstein equation using the background
metric $\delta_{i j}$, that is, we operate on Eq.\
(\ref{eq:einstein-2ndorder}) with $\p_i = \delta_{i j} \p^j$ (recall that
$\p_i \chi^{i \ixt}=0$) together with the inverse of the spatial Laplacian,
denoted by $\triangle^{-1}$.  The result is
\begin{equation} \label{eq:dphi2dsigma2}
  \frac{1}{2} \left( \vp'_0 \,\ddphi + \sigma'_0 \,\ddsigma \right) =
  \frac{{\psi^{\ixt}}' + \Hcal \phi^{\ixt} + \triangle^{-1} \alpha}{\kappa^2}
  - \triangle^{-1} \beta,
\end{equation}
where
\begin{align}
  \alpha =\, & 2\, {\phi^{\ixo}}' \p_i \p^i \phi^{\ixo} + 10\, \p_i
  {\phi^{\ixo}}' \p^i \phi^{\ixo} + 8\, \phi^{\ixo} \p_i \p^i {\phi^{\ixo}}',
  \nn\\
  \beta =\, & \p_i \dphi' \p^i \dphi + \p_i \dsigma' \p^i \dsigma + \dphi'
  \p_i
  \p^i \dphi + \dsigma' \p_i \p^i \dsigma  \label{eq:alphajabeta} \\
  & + 2\, \phi^{\ixo} \p_i \p^i \left( \vp'_0 \dphi + \sigma'_0 \dsigma
  \right) + 2\, \p_i \phi^{\ixo} \p^i \left( \vp'_0 \dphi + \sigma'_0 \dsigma
  \right). \nn
\end{align}

Taking the trace of the $ij$ component of the second order Einstein equation
and recalling that $\chi^{i \ixt}_{\phantom i i} = \p^i \chi^{\ixt}_i = 0$ we
obtain
\begin{align}
  \frac{1}{3} \p_i \p^i \phi^{\ixt} - \frac{1}{3} \p_i \p^i \psi^{\ixt} &= -
  \Hcal {\phi^{\ixt}}' - \frac{a''}{a} \phi^{\ixt} - \Hcal^2 \phi^{\ixt} -
  {\psi^{\ixt}}'' - 2 \Hcal {\psi^{\ixt}}' + 8 \frac{a''}{a} \left(
    \phi^{\ixo} \right)^2
  \nn\\
  & - 4 \Hcal^2 \left( \phi^{\ixo} \right)^2 + 8 \Hcal \phi^{\ixo}
  {\phi^{\ixo}}' + \frac{7}{3} \p_i \phi^{\ixo} \p^i \phi^{\ixo} + \frac{8}{3}
  \phi^{\ixo} \p_i \p^i \phi^{\ixo}
  \nn\\
  & + \left( {\phi^{\ixo}}' \right)^2 + \kappa^2 \left\{ \frac{1}{2} \left(
      \vp'_0 \,\ddphi' + \sigma'_0 \,\ddsigma' \right) + \frac{1}{2} \left(
      \left( \dphi' \right)^2 + \left( \dsigma' \right)^2 \right) \right.
  \nn\\
  & - 2 \left( \vp'_0 \,\dphi' + \sigma'_0 \,\dsigma' \right) \phi^{\ixo} + 2
  w^2 \left( \phi^{\ixo} \right)^2
  \nn\\
  & - \,\frac{1}{6} \left( \p_i \dphi \,\p^i \dphi + \p_i \dsigma \,\p^i
    \dsigma \right) - \frac{a^2}{2} \left[ \frac{\p V}{\p \vp} \ddphi +
    \frac{\p V}{\p \sigma} \ddsigma \right.
  \nn\\
  & \left. \left. {}-\frac{\p^2 V}{\p \vp^2} \left( \dphi \right)^2 +
  \frac{\p^2 V}{\p \sigma^2} \left( \dsigma \right)^2 + 2 \,\frac{\p^2 V}{\p
  \vp \,\p \sigma} \dphi \,\dsigma \right] \right\}~.
  \label{eq:phi2psi2-1}
\end{align}
From Eq.\ (\ref{eq:dphi2dsigma2}) and its derivative with respect
to time we obtain an expression for the second order terms of the
metric perturbations on the right hand side, which transforms the
Eq.\ (\ref{eq:phi2psi2-1}) into
\begin{align}
  \frac{1}{3} \p_i \p^i \phi^{\ixt} - \frac{1}{3} \p_i \p^i \psi^{\ixt} &= 8
  \frac{a''}{a} \left( \phi^{\ixo} \right)^2 - 4 \Hcal^2 \left( \phi^{\ixo}
  \right)^2 + 8 \Hcal \phi^{\ixo} {\phi^{\ixo}}' + \frac{7}{3} \p_i
  \phi^{\ixo} \p^i \phi^{\ixo}
  \nonumber\\
  & + \frac{8}{3} \phi^{\ixo} \p_i \p^i \phi^{\ixo} + \left( {\phi^{\ixo}}'
  \right)^2 + \triangle^{-1} \alpha' + 2 \Hcal \triangle^{-1} \alpha -
  \kappa^2 \triangle^{-1} \beta'
  \nonumber\\
  & - 2 \Hcal \kappa^2 \triangle^{-1} \beta + \kappa^2 \left\{\frac{1}{2}
    \left( \left( \dphi' \right)^2 + \left( \dsigma' \right)^2 \right) \right.
  \nonumber\\
  & - 2 \left( \vp'_0 \,\dphi' + \sigma'_0 \,\dsigma' \right) \phi^{\ixo} + 2
  w^2 \left( \phi^{\ixo} \right)^2
  \nonumber\\
  & - \,\frac{1}{6} \left( \p_i \dphi \,\p^i \dphi + \p_i \dsigma \,\p^i
    \dsigma \right)
  \nonumber\\
  & \left. - \frac{a^2}{2} \left[ \frac{\p^2 V}{\p \vp^2} \left( \dphi
      \right)^2 + \frac{\p^2 V}{\p \sigma^2} \left( \dsigma \right)^2 + 2
      \,\frac{\p^2 V}{\p \vp \,\p \sigma} \dphi \,\dsigma \right] \right\}~.
  \label{eq:phi2psi2-2}
\end{align}

Taking the inverse Laplacian of the previous equation (\ref{eq:phi2psi2-2})
leads to
\begin{equation}
  \psi^{\ixt} = \phi^{\ixt} - \triangle^{-1} \gamma~,\label{eq:gammaimplicit}
\end{equation}
where $\gamma$ is three times the right hand side of Eq.\
(\ref{eq:phi2psi2-2}). We then plug this result into Eq.\
(\ref{eq:dphi2dsigma2}) and obtain
\begin{equation} \label{eq:dphi2dsigma2-2}
  \frac{1}{2} \left( \vp'_0 \,\ddphi + \sigma'_0 \,\ddsigma \right) =
  \frac{{\phi^{\ixt}}' + \Hcal \phi^{\ixt} + \triangle^{-1} \alpha}{\kappa^2}
  - \triangle^{-1} \beta - \frac{1}{\kappa^2} \triangle^{-1} \gamma'~.
\end{equation}

Thus far we have made use of the $i0$ and $ij$ components of the second order
Einstein equations. By utilizing Eq.\ (\ref{eq:gammaimplicit}) together with
the derivative of Eq.\ (\ref{eq:dphi2dsigma2-2}) with respect to the conformal
time $\tau$ the $00$ component of the second order Einstein equation can be
written as
\begin{align}
  & {\phi^{\ixt}}'' - \p_i \p^i \phi^{\ixt} + 2 \Hcal {\phi^{\ixt}}' + 2
  \Hcal' \phi^{\ixt} = 12 \Hcal^2 \left( \phi^{\ixo} \right)^2 + 3 \left(
    {\phi^{\ixo}}' \right)^2 + 8 \phi^{\ixo} \p_i \p^i \phi^{\ixo}
  \nonumber\\
  & + 3 \p_i \phi^{\ixo} \p^i \phi^{\ixo} + 2 \Hcal \triangle^{-1} \alpha - 2
  \Hcal \kappa^2 \triangle^{-1} \beta - 2 \Hcal \triangle^{-1} \gamma' -
  \triangle^{-1} \alpha' + \kappa^2 \triangle^{-1} \beta'
  \nonumber\\
  & + \triangle^{-1} \gamma'' + 3 \Hcal \triangle^{-1} \gamma' - \gamma +
  \kappa^2 \left\{ \vp''_0 \,\ddphi + \sigma''_0 \,\ddsigma - \frac{1}{2}
    \left[ \left( \dphi' \right)^2 + \left( \dsigma' \right)^2 \right] \right.
  \nonumber\\
  & - \frac{1}{2} \left( \p_i \dphi \,\p^i \dphi + \p_i \dsigma \,\p^i \dsigma
  \right) - 2 w^2 \left( \phi^{\ixo}
  \right)^2 + 2 \left( \vp'_0 \dphi' + \sigma'_0 \dsigma' \right) \phi^{\ixo}
  \nonumber\\
  & \left. - \frac{a^2}{2} \left[ \frac{\p^2 V}{\p \vp^2} \left( \dphi
      \right)^2 + \frac{\p^2 V}{\p \sigma^2} \left( \dsigma \right)^2 + 2
      \,\frac{\p^2 V}{\p \vp \,\p \sigma} \dphi \,\dsigma \right] \right\}~,
      \label{eq:phi2master}
\end{align}
where we have also applied the background field equations \cite{liddle00}
$\vp''_0 + 2 \Hcal \vp'_0 + a^2 \p V / \p \vp = 0$ and $\sigma''_0 + 2 \Hcal
\sigma'_0 + a^2 \p V / \p \sigma = 0$.

Eq.\ (\ref{eq:phi2master}) is our master equation describing the evolution of
the second order metric perturbation $\phi^{\ixt}$.  We have not made any
approximations in deriving it; therefore it applies generally to all
inflationary models with two scalar fields. Understanding the evolution of
$\phi^{\ixt}$ is of key importance since it is going to be used later in
calculating the second order gauge invariant curvature perturbation
$\Rcal^{\ixt}$.

\section{Hybrid inflation}

\subsection{Basic features of the model}

Until now our treatment has been very general and we have not made any
assumptions, e.g., about the form of the potential. In the second order we now
have three different equations but four unknown functions $\phi^{\ixt}$,
$\psi^{\ixt}$, $\ddphi$, and $\ddsigma$.  In fact, as is noted in
\cite{polarski94}, the dynamics of a system of two scalar fields can not be
described by just one equation even in the first order. Therefore, we need an
additional constraint in order to solve completely the dynamics of the system.

In our case such a constraint is provided by the fact that in hybrid inflation
models the second scalar field $\sigma$, the ``waterfall field'', sits at the
bottom of a non-curved valley so that we may take $\sigma_0=0$. Consistent
with this we also assume that the inflation potential does not contain any
terms linear in $\sigma$ so that $\sigma_0=0$ is indeed a local
minimum\footnote{Note that our formalism would apply also to the
  Linde--Mukhanov model \cite{linde96}.}. As we will discuss, in hybrid
inflation the perturbations in $\sigma$ are decoupled from the evolution of
$\phi^{\ixt}$ to the extent that the dynamics of $\delta\sigma$ can be
considered independently.

The potential of the hybrid inflation scenario reads \cite{lyth99a}
\begin{equation} \label{eq:hybridpotential}
  V(\varphi,\sigma)
  = V_0 - \frac{1}{2}m^2_0 \sigma^2 + \frac{1}{4} \lambda
  \sigma^4 + \frac{1}{2} m^2 \vp^2 + \frac{1}{2} g^2 \sigma^2
  \vp^2.
\end{equation}
As the parameters are related by
$V_0 = m^2_0/4\lambda$, the number of free parameters is four.

The critical value of the inflaton field defines the point where
the "valley" in the $\sigma$ direction disappears. It is given by
\begin{equation}
  \vp^2_c = \frac{m^2_0}{g^2}~.
\end{equation}
Inflation ends when $\vp$ reaches the critical value. Depending on the
potential in the $\sigma$ direction, there could be a few additional e-folds
due to the slow initial motion of $\sigma$.
The usual slow-roll parameters are defined by \cite{liddle00,riotto02}
\begin{align}
  \epsilon &\equiv \frac{1}{2 \kappa^2} \left( \frac{1}{V} \frac{\p V}{\p \vp}
  \right)^2= -\frac{\dot H}{H^2} = \frac{\kappa^2}{2}
  \frac{\dot{\vp}_0^2}{H^2}~,  \label{eq:sr-epsilon}\\
  \eta &\equiv \frac{1}{\kappa^2} \frac{1}{V} \frac{\p^2 V}{\p \vp^2}=\epsilon
  - \frac{\ddot{\vp}_0}{H \dot{\vp}_0}~,
  \label{eq:sr-eta}
\end{align}
where the dot denotes derivation with respect to the cosmic time, and $H=\dot
a/a$ is the Hubble parameter. During inflation, i.e.\ when $\vp > \vp_c$,
$\sigma \simeq 0$ and $V_0$ dominates the hybrid potential. Hence, using the
potential (\ref{eq:hybridpotential}) the slow-roll parameters read
\begin{equation}
  \epsilon = \frac{1}{2 \kappa^2} \left(\frac{m^2 \vp}{V_0} \right)^2,~~
  \eta = \frac{m^2}{\kappa^2 V_0}~.
\end{equation}
During the slow roll, the inflaton field receives perturbations with a power
law spectrum $\Pcal_{\vp} \equiv k^3 | \delta\vp |^2/2\pi \sim k^{n-1}$. The
almost scale invariant nature of the perturbations is guaranteed by the
smallness of the slow-roll parameters with $n = 1 - 6 \epsilon + 2 \eta$,
where $n=1$ corresponds to the exact scale invariance.

By virtue of the construction of the model, the perturbations
generated by the inflaton are mainly Gaussian. Terms quadratic in
$\vp$ produce perturbations proportional to $\vp_0 \,\dphi$. The
waterfall field $\sigma$ does not give rise to this kind of
perturbations since its background value $\sigma_0=0$. Therefore
the leading order perturbations are Gaussian and due to the
inflaton. However, as we will show below, $\sigma$ may well induce
second order non-Gaussian perturbations which are larger than the
small second order inflaton perturbations and could be
observationally significant.

Focusing now on hybrid inflation and setting $\sigma_0=0$ (together with $\p^2
V / \p \sigma \p \vp = 0$), Eq.\ (\ref{eq:phi2master}) can be written as
\begin{align}
  & {\phi^{\ixt}}'' - \p_i \p^i \phi^{\ixt} + 2 \Hcal {\phi^{\ixt}}' + 2
  \Hcal' \phi^{\ixt} = 12 \Hcal^2 \left( \phi^{\ixo} \right)^2 + 3 \left(
    {\phi^{\ixo}}' \right)^2 + 8 \phi^{\ixo} \p_i \p^i \phi^{\ixo}
  \nonumber\\
  & + 3 \p_i \phi^{\ixo} \p^i \phi^{\ixo} + 2 \Hcal \triangle^{-1} \alpha - 2
  \Hcal \kappa^2 \triangle^{-1} \beta  -
  \triangle^{-1} \alpha' + \kappa^2 \triangle^{-1} \beta'
  \nonumber\\
  & + \triangle^{-1} \gamma'' + \Hcal \triangle^{-1} \gamma' - \gamma +
  \kappa^2 \left\{ \vp''_0 \,\ddphi - \frac{1}{2} \left[ \left( \dphi'
      \right)^2 + \left( \dsigma' \right)^2\right] \right.
  \nonumber\\
  & - \frac{1}{2} \left( \p_i \dphi \,\p^i \dphi + \p_i \dsigma \,\p^i \dsigma
  \right) - 2 {\vp_0'}^2 \left( \phi^{\ixo} \right)^2 + 2 \vp'_0 \dphi'
  \phi^{\ixo}
  \nonumber\\
  & \left. - \frac{a^2}{2} \left[ \frac{\p^2 V}{\p \vp^2} \left( \dphi
      \right)^2 + \frac{\p^2 V}{\p \sigma^2} \left( \dsigma \right)^2 \right]
  \right\}~,
  \label{eq:phi2master-2}
\end{align}
where the definition of $\alpha$ does not change but $\beta$ and $\gamma$
simplify somewhat with respect to their original forms (see Eqs.\
(\ref{eq:alphajabeta}), (\ref{eq:gammaimplicit})). From Eq.\
(\ref{eq:dphi2dsigma2-2}) we obtain an expression for $\ddphi$ which we then
plug into (\ref{eq:phi2master-2}). The result is a master equation for the
second order metric perturbation $\phi^{\ixt}$ in hybrid inflation:
\begin{align}
  & {\phi^{\ixt}}'' - \p_i \p^i \phi^{\ixt} + 2 \left( \Hcal -
    \frac{\vp''_0}{\vp'_0} \right) {\phi^{\ixt}}' + 2 \left( \Hcal' -
    \frac{\vp''_0}{\vp'_0} \Hcal \right) \phi^{\ixt} = \nonumber\\
  & 12 \Hcal^2 \left( \phi^{\ixo} \right)^2 + 3 \left( {\phi^{\ixo}}'
  \right)^2 + 8 \phi^{\ixo} \p_i \p^i \phi^{\ixo} + 3 \p_i \phi^{\ixo} \,\p^i
  \phi^{\ixo} + 2 \left( \Hcal + \frac{\vp''_0}{\vp'_0} \right) \triangle^{-1}
  \alpha \nonumber\\
  & - \triangle^{-1} \alpha' - 2 \kappa^2 \left( \Hcal +
    \frac{\vp''_0}{\vp'_0} \right) \triangle^{-1} \beta + \kappa^2
  \triangle^{-1} \beta' - \gamma + \left( \Hcal - 2 \frac{\vp''_0}{\vp'_0}
  \right) \triangle^{-1} \gamma' + \triangle^{-1} \gamma'' \nonumber\\
  & + \kappa^2 \left\{ -\frac{1}{2} \left[ \left( \dphi' \right)^2 + \left(
        \dsigma' \right)^2 \right] - \frac{1}{2} \left( \p_i \dphi \p^i \dphi
      + \p_i \dsigma \p^i \dsigma \right) -2 {\vp'_0}^2 \left( \phi^{\ixo}
    \right)^2
  \right. \nonumber\\
  & \left. + 2 \vp'_0 \phi^{\ixo} \dphi' - \frac{a^2}{2} \left[ \frac{\p^2
        V}{\p \vp^2} \left( \dphi \right)^2 + \frac{\p^2 V}{\p \sigma^2}
      \left( \dsigma \right)^2  \right] \right\}.\label{eq:hybridmaster}
\end{align}
The important point is that now all the source terms are quadratic in the
first order perturbations. Note that in Eq.\ (\ref{eq:hybridmaster}) we have
not made any approximations.

\subsection{Metric perturbation in hybrid inflation}

It is convenient to rewrite our master equation Eq.\ (\ref{eq:hybridmaster})
in terms of cosmic time $dt=a d\tau$. This makes it easier to make use of the
slow-roll parameters (\ref{eq:sr-epsilon}) and (\ref{eq:sr-eta}). We find
\begin{align}
  & {\ddot{\phi}}^{\ixt} + H \left( 1 - 2 \frac{\ddot{\vp}_0}{H \dot{\vp}_0}
  \right) {\dot{\phi}}^{\ixt} + 2 H^2 \left( \frac{\dot H}{H^2} -
    \frac{\ddot{\vp}_0}{H \dot{\vp}_0} \right) \phi^{\ixt} - \frac{1}{a^2}
  \p_i \p^i \phi^{\ixt}
  \nonumber\\
  & = -24 H^2 \left( 1 + \frac{\dot H}{H^2} \right) \left( \phi^{\ixo}
  \right)^2 - 24 H \phi^{\ixo} \dot{\phi}^{\ixo} - \frac{4}{a^2} \p_i
  \phi^{\ixo} \,\p^i \phi^{\ixo}
  \nonumber\\
  & \mbox{} + 2 H \left( 3 - \frac{\ddot{\vp}_0}{H \dot{\vp}_0} \right)
  \triangle^{-1} \left( \kappa^2 \frac{\beta}{a} - \frac{\alpha}{a} \right) +
  4 \triangle^{-1} \left( \kappa^2 \frac{\beta}{a} - \frac{\alpha}{a}
  \right)^{\largedot}
  \nonumber\\
  & \mbox{} -2 \frac{\ddot{\vp}_0}{\dot{\vp}_0} \triangle^{-1} \dot{\gamma} +
  \triangle^{-1} \ddot{\gamma}
  \nonumber\\
  & \mbox{} - \kappa^2 \left\{ 2 \left[ \left( \dphid \right)^2 + \left(
        \dsigmad \right)^2 \right] + 8 \dot{\vp}_0^2 \left( \phi^{\ixo}
    \right)^2 - 8 \dot{\vp}_0 \phi^{\ixo} \dphid \phantom{\frac{1}{2}} \right.
  \nonumber\\
  & \mbox{} \left. - \left[ \frac{\p^2 V}{\p \vp^2} \left( \dphi \right)^2 +
      \frac{\p^2 V}{\p \sigma^2} \left( \dsigma \right)^2 \right] \right\}~,
  \label{eq:phi2master-3}
\end{align}
where we have written $-\gamma$ explicitly (see Eq.\ (\ref{eq:gammaimplicit})).

We need to evaluate the term $\triangle^{-1} \left( \kappa^2 \beta / a -
  \alpha / a \right)$ in Eq.\ (\ref{eq:phi2master-3}). In terms of the cosmic
time $t$ we may simply write
\begin{equation}
  \frac{\alpha}{a} = 2 \,\p_i \p^i \left( \phi^{\ixo} \dot{\phi}^{\ixo} \right)
  + 6 \left( \p_i \phi^{\ixo} \p^i \dot{\phi}^{\ixo} + \phi^{\ixo} \p_i \p^i
  \dot{\phi}^{\ixo} \right)
\end{equation}
(see Eq.\ (\ref{eq:alphajabeta})). To derive an expression for
$\beta/a$ we make use of the first order Einstein equation
(\ref{eq:einstein-1storder}). Setting $\psi^{\ixo} = \phi^{\ixo}$
we obtain $\dphid$ from the $00$ component and from the $0i$
component an equation of motion for the first order metric
perturbation given by
\begin{equation} \label{eq:phieom}
  \dot{\phi}^{\ixo} + H \phi^{\ixo} = \frac{\kappa^2}{2} \,\dot{\vp}_0
  \,\dphi~.
\end{equation}
After some algebra we can write
\begin{align}
  \frac{\beta}{a} &= \frac{1}{2} \frac{\ddot{\vp}_0}{\dot{\vp}_0} \,\p_i \p^i
  \left( \dphi \right)^2 + 3 \,\dot{\vp}_0 \,\phi^{\ixo} \,\p_i \p^i \dphi + 3
  \,\dot{\vp}_0 \,\p_i \phi^{\ixo} \,\p^i \dphi \nn\\
  &\mbox{} + \frac{2}{\kappa^2 a^2 \dot{\vp}_0} \,\p_i \p_k \p^k \phi^{\ixo}
  \,\p^i \dphi + \frac{2}{\kappa^2 a^2 \dot{\vp}_0} \,\p_k \p^k \phi^{\ixo}
  \,\p_i \p^i \dphi \nn\\
  &\mbox{} + \p_i \dsigma \,\p^i \dsigmad + \dsigmad \,\p_i \p^i
  \dsigma~. \phantom{\frac{1}{2}}
\end{align}
Thus we obtain
\begin{align}
  \triangle^{-1} \left( \kappa^2 \frac{\beta}{a} - \frac{\alpha}{a} \right) &=
  \frac{\kappa^2}{2} \frac{\ddot{\vp}_0}{\dot{\vp}_0} \left( \dphi \right)^2 +
  3 H \left( \phi^{\ixo} \right)^2 - 2 \phi^{\ixo} \dot{\phi}^{\ixo}
  \nn\\
  &\mbox{} + \frac{2}{a^2 \dot{\vp}_0} \triangle^{-1} \left( \p_i \p_k \p^k
  \phi^{\ixo} \,\p^i \dphi + \p_k \p^k \phi^{\ixo} \,\p_i \p^i \dphi \right)
  \nn\\
  &\mbox{} + \kappa^2 \triangle^{-1} \p_i \left( \dsigmad \,\p^i \dsigma
  \right)~, \phantom{\frac{1}{2}} \label{eq:triangleterm}
\end{align}
where we have again utilized the equation of motion Eq.\
(\ref{eq:phieom}).

In hybrid inflation constraint $\sigma$ decouples completely from the first
order perturbed Einstein equations Eq.\ (\ref{eq:einstein-1storder}) and we
may isolate the contributions due to $\sigma$ from those that are due to the
inflaton $\vp$.  Taking into account Eqs.\ (\ref{eq:alphajabeta}),
(\ref{eq:phi2psi2-2}), (\ref{eq:gammaimplicit}) and (\ref{eq:triangleterm}) we
thus write
\begin{align} \label{eq:abcsplit}
  \alpha &\equiv \alpha_{\vp}, \nn\\
  \beta &\equiv \beta_{\vp} + \beta_{\sigma} = \beta_{\vp} + a \,\p_i \left(
    \dsigmad \p^i \dsigma \right), \\
  \gamma &\equiv \gamma_{\vp} + \gamma_{\sigma} = \gamma_{\vp} - 9 a^2
  \kappa^2 H \triangle^{-1} \p_i \left( \dsigmad \p^i \dsigma \right) - 3 a^2
  \kappa^2 \triangle^{-1} \p_i \left( \dsigmad \p^i \dsigma
  \right)^{\largedot}
  \nn\\
  & \phantom{1+1+1+} + a^2 \kappa^2 \left[ \frac{3}{2} \left( \dsigmad
    \right)^2 - \frac{1}{2 a^2} \p_i \dsigma \,\p^i \dsigma - \frac{3}{2}
    \frac{\p^2 V}{\p \sigma^2} \left( \dsigma \right)^2 \right], \nn
\end{align}
and
\begin{align}
  \triangle^{-1} \left( \kappa^2 \frac{\beta}{a} - \frac{\alpha}{a} \right)
  &\equiv \triangle^{-1} \left( \kappa^2 \frac{\beta}{a} - \frac{\alpha}{a}
  \right)_{\vp} + \triangle^{-1} \left( \kappa^2 \frac{\beta}{a} -
    \frac{\alpha}{a} \right)_{\sigma} \nonumber\\
  &= \triangle^{-1} \left( \kappa^2 \frac{\beta}{a} - \frac{\alpha}{a}
  \right)_{\vp} + \kappa^2 \triangle^{-1} \p_i \left( \dsigmad \p^i \dsigma
  \right), \label{eq:trianglesplit}
\end{align}
where the subscripts $\vp$ and $\sigma$ denote respectively the inflaton and
$\sigma$ contributions. The inflaton contributions have all been computed
previously by Acquaviva et al.\ \cite{acquaviva03}, whose treatment is still
valid here due to the decoupling of $\vp$ and $\sigma$. Plugging Eq.\
(\ref{eq:trianglesplit}) into Eq.\ (\ref{eq:phi2master-3}) and dropping terms
next to the leading order in slow-roll parameters the master equation Eq.\
(\ref{eq:phi2master-3}) can be written as
\begin{align}
  & {\ddot{\phi}}^{\ixt} + H {\dot{\phi}}^{\ixt} + 2 H^2 \left( \frac{\dot
      H}{H^2} - \frac{\ddot{\vp}_0}{H \dot{\vp}_0} \right) \phi^{\ixt} -
  \frac{1}{a^2} \p_i \p^i \phi^{\ixt}
  \nonumber\\
  & = \left[~\textrm{inflaton contribution}~\right] \phantom{\frac{1}{2}}
  \nonumber\\
  & + 6 \kappa^2 H \triangle^{-1} \p_i \left( \dsigmad \,\p^i \dsigma \right)
  + 4 \kappa^2 \triangle^{-1} \p_i \left( \dsigmad \,\p^i \dsigma
  \right)^{\largedot} \phantom{\frac{1}{2}}
  \nonumber\\
  & - 2 \kappa^2 \left( \dsigmad \right)^2 + \kappa^2 \frac{\p^2 V}{\p
    \sigma^2} \left( \dsigma \right)^2 - 2 \frac{\ddot{\vp}_0}{\dot{\vp}_0}
  \triangle^{-1} \dot{\gamma}_{\sigma} + \triangle^{-1}
  \ddot{\gamma}_{\sigma}~. \label{eq:mastersplit}
\end{align}
The inflaton part can be computed from Eq.\ (\ref{eq:phi2master-3}) but it has
already been done in \cite{acquaviva03}. The last two lines in Eq.\
(\ref{eq:mastersplit}) represent the second order metric perturbation solely
due to the second field of hybrid inflation.

We have now succeeded in completely isolating the contribution of
$\sigma$ to the dynamics of the second order metric perturbation
$\phi^{\ixt}$ within the hybrid inflation paradigm. This is of
utmost importance since the master equation, Eq.\
(\ref{eq:mastersplit}), plays a key role when we later compute the
curvature perturbation. The decoupling of $\vp$ and $\sigma$ makes
it also possible to isolate the contribution of $\sigma$ to the
curvature perturbation also.

\section{Curvature perturbations}

\subsection{General formula}

Let us now derive the second-order gauge-invariant curvature perturbation for
two scalar fields following the procedure described in \cite{acquaviva03} (for
an exact treatment see \cite{malik03}). We denote the gauge-invariant
curvature perturbation by $\Rcal$ and expand it up to the second order in the
already familiar way\footnote{Recall that the curvature perturbation does not
  have a homogeneous part.}
\begin{equation}
  \Rcal = \Rcal^{\ixo} + \frac{1}{2} \Rcal^{\ixt}~.
\end{equation}
We are mainly interested in the second order part. Our starting
point is the first order quantity \cite{gordon01}
\begin{equation}
  \Rcal^{\ixo} = \psi^{\ixo} + \Hcal \left(\frac{\vp'_0 \dphi + \sigma'_0
  \dsigma}{ {\vp'_0}^2 + {\sigma'_0}^2 }  \right)~.
\end{equation}
Instead of the first order quantities we write the expansion up to second
order for the metric perturbation $\psi = \psi^{\ixo} + \frac{1}{2}
\psi^{\ixt}$ and the scalar fields $\delta \vp = \dphi + \frac{1}{2} \ddphi$
and $\delta \sigma = \dsigma + \frac{1}{2} \ddsigma$.
We obtain
\begin{equation}
  \psi + \Hcal \left( \frac{\vp'_0 \,\delta \vp + \sigma'_0 \,\delta \sigma}{
  {\vp'_0}^2 + {\sigma'_0}^2 } \right) = \Rcal^{\ixo} + \frac{1}{2}  \left[
  \psi^{\ixt} + \Hcal \left( \frac{\vp'_0 \,\ddphi + \sigma'_0 \,\ddsigma}{
  {\vp'_0}^2 + {\sigma'_0}^2 } \right) \right]~. \label{eq:curvpert}
\end{equation}

Consider then the following second-order shift of the time coordinate
\cite{bruni97,matarrese98,acquaviva03}
\begin{equation}
  \tau \to \tau - \xi^0_{\ix 1} + \frac{1}{2} \left( {\xi^0_{\ixo}}'
    \xi^0_{\ixo} - \xi^0_{\ixt} \right)~, \label{muunnos}
\end{equation}
which transforms $\psi^{\ixt}$, $\ddphi$ and $\ddsigma$ into
\begin{eqnarray}
  \widetilde{\psi^{\ixt}} &=& \psi^{\ixt} + 2 \xi^0_{\ixo} \left(
  {\psi^{\ixo}}' + 2 \Hcal \psi^{\ixo} \right) - \left( \Hcal' + 2 \Hcal^2
  \right) \left( \xi^0_{\ixo} \right)^2 \nonumber\\
  && \mbox{} - \Hcal {\xi^0_{\ixo}}' \xi^0_{\ixo} - \Hcal \xi^0_{\ixt}
  - \frac{1}{3} \left( 2 \p^i \omega^{\ixo} - \p^i \xi^0_{\ixo} \right)
  \p_i \xi^0_{\ixo}~, \\ \nn
  \widetilde{\ddphi} &=& \ddphi + \xi^0_{\ixo} \left( \vp''_0 \xi^0_{\ixo} +
  \vp'_0 {\xi^0_{\ixo}}' + 2 \dphi' \right) + \vp'_0 \xi^0_{\ixt}~, \\ \nn
  \widetilde{\ddsigma} &=& \ddsigma + \xi^0_{\ixo} \left( \sigma''_0
  \xi^0_{\ixo} + \sigma'_0 {\xi^0_{\ixo}}' + 2 \dsigma' \right) + \sigma'_0
  \xi^0_{\ixt}~.
\end{eqnarray}
Thus, the expansion (\ref{eq:curvpert}) transforms as\footnote{The first order
  part $\Rcal^{\ixo}$ remains unchanged under the transformation Eq.\ 
  (\ref{muunnos}).}
\begin{eqnarray}
  \widetilde{\psi} &+& \Hcal \left( \frac{\vp'_0 \,\widetilde{\delta \vp} +
  \sigma'_0 \,\widetilde{\delta \sigma}}{ {\vp'_0}^2 + {\sigma'_0}^2 } \right)
  \nonumber\\
  &=& \Rcal^{\ixo} + \frac{1}{2}  \left[ \Hcal \left( \frac{\vp'_0
  \,\widetilde{\ddphi} + \sigma'_0 \,\widetilde{\ddsigma}}{ {\vp'_0}^2 +
  {\sigma'_0}^2 } \right)  + \widetilde{\psi^{\ixt}}  \right] \nonumber\\
  &=& \psi + \Hcal \left( \frac{\vp'_0 \,\delta \vp + \sigma'_0 \,\delta
  \sigma}{ {\vp'_0}^2 + {\sigma'_0}^2 } \right) + \xi^0_{\ixo} T \nonumber\\
  && - \frac{1}{2} \left( \xi^0_{\ixo} \right)^2 \left[ \Hcal' + 2 \Hcal^2 -
  \Hcal \left( \frac{\vp'_0 \vp''_0 + \sigma'_0 \sigma''_0}{{\vp'_0}^2 +
  {\sigma'_0}^2} \right) \right] - \frac{1}{6} \left( 2 \p^i \omega^{\ixo} -
  \p^i \xi^0_{\ixo} \right) \p_i \xi^0_{\ixo}~, \label{eq:curvexp}
\end{eqnarray}
where, following \cite{acquaviva03}, we have denoted
\begin{equation}
  T = {\psi^{\ixo}}' + 2 \Hcal \psi^{\ixo} + \Hcal \left( \frac{\vp'_0
  \,\dphi' + \sigma'_0 \,\dsigma'}{{\vp'_0}^2 + {\sigma'_0}^2} \right)~.
\end{equation}

By virtue of the first order transformations
$\widetilde{\psi^{\ixo}} = \psi^{\ixo} - \Hcal \,\xi^0_{\ixo},~
\widetilde{\dphi} = \dphi + \vp'_0 \,\xi^0_{\ixo}$ and
$\widetilde{\dsigma} =\dsigma + \sigma'_0 \,\xi^0_{\ixo}$ we find
\begin{equation}
  T - \widetilde{T} = \xi^0_{\ixo} \left[ \Hcal' + 2 \Hcal^2 - \Hcal \left(
  \frac{\vp'_0 \vp''_0 + \sigma'_0 \sigma''_0}{{\vp'_0}^2 + {\sigma'_0}^2}
  \right) \right]~. \label{eq:T-transform}
\end{equation}
The expansion (\ref{eq:curvexp}) can now be written as
\begin{eqnarray}
  \widetilde{\psi} + \Hcal \left( \frac{\vp'_0 \,\widetilde{\delta \vp} +
  \sigma'_0 \,\widetilde{\delta \sigma}}{ {\vp'_0}^2 + {\sigma'_0}^2 } \right)
  &=& \psi + \Hcal \left( \frac{\vp'_0 \,\delta \vp + \sigma'_0 \,\delta
  \sigma}{ {\vp'_0}^2 + {\sigma'_0}^2 } \right) \nonumber\\
  && + \frac{1}{2} \left( T +
  \widetilde T \right) \xi^0_{\ixo} - \frac{1}{6}\left( 2 \p^i \omega^{\ixo} -
  \p^i \xi^0_{\ixo} \right) \p_i \xi^0_{\ixo}~.
\end{eqnarray}

We also solve $\xi^0_{\ixo}$ from Eq.\ (\ref{eq:T-transform}) and insert it
into the $T + \widetilde T$ term above. Note that by virtue of the first order
transformation $\widetilde{\omega^{\ixo}} = \omega^{\ixo} - \xi^0_{\ixo}$, the
last term can be written as
\begin{equation}
  \mbox{} - \frac{1}{6} \left( 2 \p^i \omega^{\ixo} - \p^i \xi^0_{\ixo}
  \right) \p_i \xi^0_{\ixo} = \mbox{} - \frac{1}{6} \left( \p^i \omega^{\ixo}
  \,\p_i \omega^{\ixo} - \p^i \widetilde{\omega^{\ixo}} \,\p_i
  \widetilde{\omega^{\ixo}} \right)~.
\end{equation}
Therefore, after some algebra we see that
\begin{eqnarray}
  \widetilde{\psi} + \Hcal \left( \frac{\vp'_0 \,\widetilde{\delta \vp} +
  \sigma'_0 \,\widetilde{\delta \sigma}}{ {\vp'_0}^2 + {\sigma'_0}^2 } \right)
  + \frac{1}{2} \,\frac{\widetilde{T}^2}{\Hcal' + 2 \Hcal^2 - \Hcal \left(
  \frac{\vp'_0 \vp''_0 + \sigma'_0 \sigma''_0}{{\vp'_0}^2 + {\sigma'_0}^2}
  \right) } - \frac{1}{6} \p^i \widetilde{\omega^{\ixo}} \,\p_i
  \widetilde{\omega^{\ixo}}\phantom{~.}&& \nonumber \\
  = \psi + \Hcal \left( \frac{\vp'_0 \,\delta \vp + \sigma'_0 \,\delta
  \sigma}{ {\vp'_0}^2 + {\sigma'_0}^2 } \right) + \frac{1}{2}
  \,\frac{T^2}{\Hcal' + 2 \Hcal^2 - \Hcal \left(
  \frac{\vp'_0 \vp''_0 + \sigma'_0 \sigma''_0}{{\vp'_0}^2 + {\sigma'_0}^2}
  \right) } - \frac{1}{6} \p^i \omega^{\ixo} \,\p_i \omega^{\ixo}~.&&
\end{eqnarray}

The treatment above shows that the comoving curvature perturbation $\Rcal =
\Rcal^{\ixo} + \frac{1}{2} \Rcal^{\ixt}$, which is invariant under the time
shift $\tau \to \tau - \xi^0_{\ix 1} + \frac{1}{2} \left( {\xi^0_{\ixo}}'
  \xi^0_{\ixo} - \xi^0_{\ixt} \right)$, reads in the case of two scalar fields
as
\begin{eqnarray}
  \Rcal &=& \Rcal^{\ixo} + \frac{1}{2} \left( \Hcal \,\frac{\vp'_0 \,\ddphi +
  \sigma'_0 \,\ddsigma}{{\vp'_0}^2 + {\sigma'_0}^2} + \psi^{\ixt} \right)
  \nonumber\\
  && \mbox{} + \frac{1}{2} \frac{\left( {\psi^{\ixo}}' + 2 \, \Hcal
  \psi^{\ixo} + \Hcal \,\frac{\vp'_0 \,\dphi' + \sigma'_0
  \,\dsigma'}{{\vp'_0}^2 + {\sigma'_0}^2} \right)^2}{\Hcal' + 2 \Hcal^2 -
  \Hcal \,\frac{\vp'_0 \,\vp''_0 + \sigma'_0 \,\sigma''_0}{{\vp'_0}^2 +
  {\sigma'_0}^2}}
  \nonumber\\
  && \mbox{} - \frac{1}{6} \p^i \omega^{\ixo} \,\p_i \omega^{\ixo}~,
\end{eqnarray}
where
\begin{equation}
  \Rcal^{\ixo} = \psi^{\ixo} + \Hcal \,\frac{\vp'_0 \,\dphi + \sigma'_0
  \,\dsigma}{{\vp'_0}^2 + {\sigma'_0}^2}~.
\end{equation}
This result coincides with the one obtained in \cite{rigopoulos02}
once one takes into account the field redefinitions there.

\subsection{Second order curvature perturbation in hybrid inflation}

With the general longitudinal gauge, here essentially $\omega^{\ixo} = 0$, and
the hybrid inflation condition $\sigma_0 = 0$ the curvature perturbation
acquires the same functional form as in the single-field case
\cite{acquaviva03}
\begin{equation}
  \Rcal = \Rcal^{\ixo} + \frac{1}{2} \left( \Hcal \frac{\ddphi}{\vp'_0} +
  \psi^{\ixt} \right) + \frac{1}{2} \frac{\left({\psi^{\ixo}}' + 2 \, \Hcal
  \psi^{\ixo} + \Hcal \,\dphi' / \vp'_0 \right)^2}{\Hcal' + 2 \,\Hcal^2 -
  \Hcal \,\vp''_0 / \vp'_0}~,
\end{equation}
where
\begin{equation}
  \Rcal^{\ixo} = \psi^{\ixo} + \Hcal \,\frac{\dphi}{\vp'_0}~.
\end{equation}\\
Rewriting $\Rcal^{\ixt}$ in terms of the cosmic time $dt=a\,d\tau$ and
applying the condition $\psi^{\ixo} = \phi^{\ixo}$, we obtain
\begin{equation}
  \Rcal^{\ixt} = H \frac{\ddphi}{\dot{\vp}_0} + \psi^{\ixt} + \frac{\left(
  \dot{\phi}^{\ixo} + 2 H \phi^{\ixo} + H \,\dphid / \dot{\vp}_0
  \right)^2}{H^2 \left( 2 + \dot H / H - \ddot{\vp}_0 / H \dot{\vp}_0
  \right)}~.
\end{equation}
Making use of the relations (\ref{eq:gammaimplicit}) and
(\ref{eq:dphi2dsigma2-2}) we find
\begin{align}
  \Rcal^{\ixt} &= \,\frac{2 H}{\kappa^2 {\dot{\vp}_0}^2} \left[
  \dot{\phi}^{\ixt} + H \phi^{\ixt} - \triangle^{-1} \left( \kappa^2
  \frac{\beta}{a} - \frac{\alpha}{a} \right) \right] + \phi^{\ixt} -
  \frac{2 H}{\kappa^2 {\dot{\vp}_0}^2} \triangle^{-1} \dot{\gamma} -
  \triangle^{-1} \gamma
  \nonumber\\
  & \mbox{} + \frac{\left( \dot{\phi}^{\ixo} + 2 H \phi^{\ixo} + H \,\dphid /
  \dot{\vp}_0 \right)^2}{H^2 \left( 2 + \dot H / H - \ddot{\vp}_0 / H
  \dot{\vp}_0 \right)}~.
\end{align}
Acquaviva et al. \cite{acquaviva03} point out that the last term gives a
subdominant contribution in the single field case. Moreover, it does not
contain any dependence on $\sigma$, not even implicitly through
$\dot{\phi}^{\ixo} + 2 H \phi^{\ixo}$. Therefore, in what follows we shall
neglect this term.

Since $2 H^2 / \kappa^2 {\dot{\vp}_0}^2 = 1 / \epsilon$, the term
$\phi^{\ixt}$ outside the square brackets is subdominant to the one inside;
hence we discard it. Thus, up to the leading order in the slow-roll parameters
we may write the curvature perturbation as
\begin{equation}
  \Rcal^{\ixt} \simeq \frac{2 H}{\kappa^2 {\dot{\vp}_0}^2} \left[
  \dot{\phi}^{\ixt} + H \phi^{\ixt} - \triangle^{-1} \left( \kappa^2
  \frac{\beta}{a} - \frac{\alpha}{a} \right) \right] - \frac{2 H}{\kappa^2
  {\dot{\vp}_0}^2} \triangle^{-1} \dot{\gamma} - \triangle^{-1} \gamma~.
  \label{eq:R2general}
\end{equation}

\subsection{Curvature perturbation due to $\sigma$}

As we did with the master equation, Eq.\ (\ref{eq:mastersplit}),
we may isolate the contributions coming from the inflaton and
$\sigma$ also in $\Rcal$. We thus write the comoving curvature
perturbation as
\begin{equation}
  \Rcal = \Rcal^{\ixo} + \frac{1}{2} \Rcal^{\ixt} = \Rcal^{\ixo}_{\vp} +
  \frac{1}{2} \Rcal^{\ixt}_{\vp} + \frac{1}{2} \Rcal^{\ixt}_{\sigma} =
  \Rcal_{\vp} + \frac{1}{2} \Rcal^{\ixt}_{\sigma}~. \label{eq:R}
\end{equation}
$\Rcal_{\vp}$ contribution has already been calculated by Acquaviva et al.
\cite{acquaviva03}. They plug the expressions (\ref{eq:alphajabeta}) and
(\ref{eq:gammaimplicit}) for $\alpha$, $\beta$, and $\gamma$ into Eq.\ 
(\ref{eq:R2general}) while $\dot{\phi}^{\ixt} + H \phi^{\ixt}$ is solved from
their master equation\footnote{This is the part marked \emph{inflaton
    contribution} in our master equation (\ref{eq:mastersplit}).}. They also
take into account the fact that at large scales, $k \ll a H$, $\psi^{\ixo}$
can be taken constant and
\begin{equation}
  \psi^{\ixo} = \frac{\kappa^2}{2} \frac{\dot{\vp}_0}{H} \dphi = \epsilon H
  \frac{\dphi}{\dot{\vp}_0}~.
\end{equation}
This makes it possible to set $\psi^{\ixo} = \epsilon \,\Rcal^{\ixo}$ so that
the result can be written in a deceptively simple looking way as
\cite{acquaviva03}
\begin{equation}
  \Rcal^{\ixt}_{\vp} = \left( \eta - 3 \epsilon \right) \left( \Rcal^{\ixo}
  \right)^2 + \Ical_{\vp}~, \label{eq:elamanvesi}
\end{equation}
where
\begin{align}
  \Ical_{\vp} = &-\frac{2}{\epsilon} \int \frac{1}{a^2} \psi^{\ixo} \p_i \p^i
  \psi^{\ixo} \,dt - \frac{4}{\epsilon} \int \frac{1}{a^2} \p_i \psi^{\ixo}
  \p^i \psi^{\ixo} \,dt \nonumber\\
  & - \frac{4}{\epsilon} \int \left( \ddot{\psi}^{\ixo} \right)^2 \,dt +
  \left( \epsilon - \eta \right) \triangle^{-1} \p_i \Rcal^{\ixo} \p^i
  \Rcal^{\ixo}~. \label{eq:Iphi}
\end{align}
The important point to stress here is that the single field contribution to
the curvature perturbation, including the integral part $\Ical_{\vp}$, is
proportional to the slow-roll parameters and hence naturally small in hybrid
inflation.

However, in hybrid inflation the waterfall field $\sigma$ yields an additional
contribution to $\Rcal$. We may calculate it from Eq.\ (\ref{eq:R2general}) by
plugging in the $\sigma$ dependent parts of $\alpha$, $\beta$, $\gamma$, and
integrating the $\sigma$ dependent part of our master equation
(\ref{eq:mastersplit}) to obtain $( \dot{\phi}^{\ixt} + H \phi^{\ixt}
)_{\sigma}$ (see \cite{acquaviva03} for the inflaton part). Leaving
$\gamma_\sigma$ implicit for the moment the result reads
\begin{align}
  \Rcal^{\ixt}_{\sigma} =\,\, &\frac{2 H}{\kappa^2 {\dot{\vp}_0}^2} \left[
    \int 6 \kappa^2 H \triangle^{-1} \p_i \left( \dsigmad \p^i \dsigma \right)
    \,dt + \int 4 \kappa^2 \triangle^{-1} \p_i \left( \dsigmad \p^i \dsigma
    \right)^{\largedot} \,dt \right.
  \nonumber\\
  & - \int 2 \kappa^2 \left( \dsigmad \right)^2 \,dt + \int \kappa^2
  \frac{\p^2 V}{\p \sigma^2} \left( \dsigma \right)^2 \,dt - \int 2
  \frac{\ddot{\vp}_0}{\dot{\vp}_0} \triangle^{-1} \dot{\gamma}_{\sigma} \,dt
  \nonumber\\
  & \left. + \int \triangle^{-1} \ddot{\gamma}_{\sigma} \,dt - \kappa^2
    \triangle^{-1} \p_i \left( \dsigmad \p^i \dsigma \right) \right] - \frac{2
    H}{\kappa^2 {\dot{\vp}_0}^2} \triangle^{-1} \dot{\gamma}_{\sigma} -
  \triangle^{-1} \gamma_{\sigma}
  \nonumber\\
  =\,\, &\frac{1}{\epsilon H} \left\{ \int \left[ \phantom{\frac{1}{2}} 6
      \kappa^2 H \triangle^{-1} \p_i \left( \dsigmad \p^i \dsigma \right) + 4
      \kappa^2 \triangle^{-1} \p_i \left( \dsigmad \p^i \dsigma
      \right)^{\largedot} \right. \right.
  \nonumber\\
  & \left. - 2 \kappa^2 \left( \dsigmad \right)^2 + \kappa^2 m^2_{\sigma}
    \left( \dsigma \right)^2 - 2 \frac{\ddot{\vp}_0}{\dot{\vp}_0}
    \triangle^{-1} \dot{\gamma}_{\sigma} + \triangle^{-1}
    \ddot{\gamma}_{\sigma} \right] \,dt
  \nonumber\\
  & \left. - \kappa^2 \triangle^{-1} \p_i \left( \dsigmad \p^i \dsigma \right)
    - \triangle^{-1} \dot{\gamma}_{\sigma} - \epsilon H \triangle^{-1}
    \gamma_{\sigma} \phantom{\int} \right\}~, \label{eq:R2sigma}
\end{align}
where we have used the shorthand notation $m^2_{\sigma} \equiv
\p^2 V / \p \sigma^2$ and the slow-roll relation Eq.\
(\ref{eq:sr-epsilon}).  Since $\gamma_\sigma$, Eq.\
(\ref{eq:abcsplit}), is not in very transparent form, we rewrite
it before substituting it into Eq.\ (\ref{eq:R2sigma}). For
instance, it is not obvious whether $\triangle^{-1}
\gamma_{\sigma}$ has terms that blow up outside the horizon. To
study this let us write
\begin{align}
  \gamma_{\sigma} = &- 9 a^2 \kappa^2 H \triangle^{-1} \p_i \left( \dsigmad
    \p^i \dsigma \right) - 3 a^2 \kappa^2 \triangle^{-1} \p_i \left( \dsigmad
    \p^i \dsigma \right)^{\largedot}
  \nonumber\\
  &+ 3 a^2 \kappa^2 \left[ \frac{1}{2} \left( \dsigmad \right)^2 - \frac{1}{6
    a^2} \p_i \dsigma \,\p^i \dsigma - \frac{1}{2} \frac{\p^2 V}{\p \sigma^2}
    \left( \dsigma \right)^2 \right]
  \nonumber\\
  = &- 3 a^2 \kappa^2 \triangle^{-1} \left[ 3 H \p_i \left( \dsigmad \p^i
    \dsigma \right) + \p_i \left( \dsigmad \p^i \dsigma \right)^{\largedot} -
    \frac{1}{2} \p_i \p^i \left( \dsigmad \right)^2 \right.
  \nonumber\\
  &+ \left. \frac{1}{6 a^2} \p_i \p^i \left( \p_k \dsigma \p^k \dsigma \right)
    + \frac{m^2_{\sigma}}{2} \p_i \p^i \left( \dsigma \right)^2 \right]
  \nonumber\\
  = &- \kappa^2 \triangle^{-1} \left[ 3 \,\p_i \left( \p_k \p^k
      \dsigma \p^i \dsigma \right) + \frac{1}{2} \p_i \p^i \left( \p_k
      \dsigma \p^k \dsigma \right) \right]~,\label{eq:gammafinal}
\end{align}
where we have used the equation of motion \cite{liddle00} $\delta^{\ixo}
\ddot{\sigma} + 3 H \dsigmad - a^{-2} \p_i \p^i \dsigma + m^2_{\sigma} \dsigma
= 0$. From the last form of Eq.\ (\ref{eq:gammafinal}) we see that
$\gamma_{\sigma}$ indeed is suppressed outside the horizon: the Fourier modes
are proportional to the wavenumber squared, $k^2$, since the inverse of the
spatial Laplacian, $\triangle^{-1}$, cancels the scale dependence caused by
two of the four spatial derivatives in each term.

Substituting Eq.\ (\ref{eq:gammafinal}) into Eq.\ (\ref{eq:R2sigma}) we can
finally spell out explicitly the contribution of $\sigma$ to $\Rcal$:
\begin{align}
  \Rcal^{\ixt}_{\sigma} =\,\, & \frac{\kappa^2}{\epsilon H} \left\{ \int
    \left[ \phantom{\frac{1}{2}} 6 H \triangle^{-1} \p_i \left( \dsigmad \p^i
        \dsigma \right) + 4 \triangle^{-1} \p_i \left( \dsigmad \p^i \dsigma
      \right)^{\largedot} \right. \right.
  \nonumber\\
  & \mbox{} - 2 \left( \dsigmad \right)^2 + m^2_{\sigma} \left( \dsigma
  \right)^2 + \left( \epsilon - \eta \right) 6 H \triangle^{-2} \p_i \left(
    \p_k \p^k \dsigma \p^i \dsigma \right)^{\largedot}
  \nonumber\\
  & \mbox{} + \left( \epsilon - \eta \right) H \triangle^{-2} \p_i \p^i \left(
    \p_k \dsigma \p^k \dsigma \right)^{\largedot} - 3
  \triangle^{-2} \p_i \left( \p_k \p^k \dsigma \p^i \dsigma
  \right)^{\largeddot}
  \nonumber\\
  & \left. \mbox{} - \frac{1}{2} \triangle^{-2} \p_i \p^i \left( \p_k \dsigma
      \p^k \dsigma \right)^{\largeddot} \right] \,dt -
    \triangle^{-1} \p_i \left( \dsigmad \p^i \dsigma \right)
    \nonumber\\
    & \mbox{} + 3 \triangle^{-2} \p_i \left( \p_k \p^k \dsigma \p^i \dsigma
    \right)^{\largedot} + \frac{1}{2} \triangle^{-2} \p_i \p^i
    \left( \p_k \dsigma \p^k \dsigma \right)^{\largedot}
    \nonumber\\
    & \left. \mbox{} + 3 \epsilon H \triangle^{-2} \p_i \left( \p_k \p^k
        \dsigma \p^i \dsigma \right) + \frac{\epsilon H}{2} \triangle^{-2}
      \p_i \p^i \left( \p_k \dsigma \p^k \dsigma \right)
    \right\}~,\label{eq:vihonviimeinen}
\end{align}
where we have used the slow-roll relations (\ref{eq:sr-epsilon}) and
(\ref{eq:sr-eta}). Various complications arising from the derivations with
respect to time and space, inverse spatial Laplacians and especially from the
integration over time still remain. The main point, however, is that we have
got rid of terms containing the metric explicitly and every term is quadratic
in the first order perturbation of $\sigma$.

\section{Large $\sigma$-induced non-Gaussianities}

An exact evaluation of the $\sigma$-induced curvature perturbation Eq.\ 
(\ref{eq:vihonviimeinen}) is beyond the scope of the present paper. However,
we can consider the orders of magnitudes and relative sizes of the various
terms appearing in Eq.\ (\ref{eq:vihonviimeinen}).  Let us first focus on the
first term outside the time integral. Since $\triangle^{-1} \p_i (\dsigmad
\p^i \dsigma ) = \triangle^{-1} \p_i \dsigmad \p^i \dsigma + \triangle^{-1}
\dsigmad \p_i \p^i \dsigma$, we see that Eq.\ (\ref{eq:vihonviimeinen})
actually contains a term of the form $(\kappa^2 / \epsilon H) \triangle^{-1}
\p_i \dsigmad \p^i \dsigma$.  Outside the horizon, where $k \ll a H$, the
first order perturbation $\dsigmad$ follows the equation of motion
$\delta^{\textrm{\tiny (1)}} \ddot{\sigma} + 3 H \dsigmad + m^2_{\sigma}
\,\dsigma = 0$. If we make the usual assumption that the mass of $\sigma$ is
small, i.e.\ $m_{\sigma} < H$, the non-decaying solution is proportional to
$\textrm{exp}(-m^2_{\sigma} t / 3 H)$. Therefore we can estimate $| \dsigmad |
\sim \frac{m^2_{\sigma}}{H} | \dsigma |$ so that
\begin{equation}
  \left| \frac{\kappa^2}{\epsilon H} \triangle^{-1} \p_i \dsigmad \,\p^i
  \dsigma \right|
  \sim \frac{m^2_{\sigma}}{H^2} \left| \triangle^{-1} \p_i \left( H
  \frac{\dsigma}{\dot{\vp}_0} \right) \,\p^i \left( H
  \frac{\dsigma}{\dot{\vp}_0} \right) \right|,
\end{equation}
where we have used the relation $\epsilon = \kappa^2 \dot{\vp}_0^2 / 2 H^2$.
The first order perturbation of an effectively massless field $\sigma$ is the
same as the first order perturbation of the inflaton field $\vp$ during
horizon exit. Hence outside the horizon we may estimate
\begin{equation}
  \left| H \frac{\dsigma}{\dot{\vp}_0} \right| \sim \left| H
  \frac{\dphi}{\dot{\vp}_0} \right| \equiv \left| \Rcal^{\ixo} \right|~.
\end{equation}
As a consequence, we finally obtain an estimate for the simplest non-integral
term in Eq.\ (\ref{eq:vihonviimeinen}):
\begin{equation}
  \left| \frac{\kappa^2}{\epsilon H} \triangle^{-1} \p_i \dsigmad \,\p^i
  \dsigma \right| \sim \frac{m^2_{\sigma}}{H^2} \left| \triangle^{-1} \p_i
  \Rcal^{\ixo} \,\p^i \Rcal^{\ixo} \right|.
\end{equation}
This result can directly be compared with the last term in $\Ical_{\vp}$ in
Eq.\ (\ref{eq:Iphi}). Because the spatial derivative operators\footnote{That
  is, the two derivatives $\p_i$ and the inverse of the spatial Laplacian
  $\triangle^{-1}$.} in $\triangle^{-1} \p_i \Rcal^{\ixo} \,\p^i \Rcal^{\ixo}$
cancel each others scale dependence, we can approximate roughly $ \left|
  \triangle^{-1} \p_i \Rcal^{\ixo} \,\p^i \Rcal^{\ixo} \right| \sim \left|
  \Rcal^{\ixo} \right|^2$. This coincides with the statement in Acquaviva et
al.\ \cite{acquaviva03} that $\Ical_{\vp}$ is of the same order in slow-roll
parameters as the contribution of the $\left( \Rcal^{\ixo} \right)^2$ term to
$\Rcal^{\ixt}_{\vp}$.

In a similar fashion we can approximate all the other terms in
$\Rcal^{\ixt}_{\sigma}$ which are not integrated over time. Inspection shows
that in every non-integrated term the spatial derivative operators cancel each
others contribution to the scale dependence. Replacing also time derivatives
with the factor $\frac{m^2_{\sigma}}{H}$ we see that there are only two kinds
of terms, those proportional to $\frac{m^2_{\sigma}}{H^2} \left| \Rcal^{\ixo}
\right|^2$ and those proportional to $\epsilon \left| \Rcal^{\ixo} \right|^2$.

The estimation of the integral part in Eq.\
(\ref{eq:vihonviimeinen}) is more involved. Because of the
integration over time the Hubble parameter $H$, slow-roll
parameters $\epsilon$ and $\eta$, mass $m_{\sigma}$ and
perturbation $\dsigma$ can no longer be assumed to be constants.
Nevertheless, we may argue that the evolution of these parameters
is relatively slow. In order to get an estimate for the integral
part without too many complications we take the time integration
to start after the scales in question have exited the horizon.  We
also assume that the quantities $H$, $\epsilon$, $\eta$,
$m_{\sigma}$ and $\dsigma$ change slowly and, for our estimation
purposes, take them to be constants. The error made this way is
not likely to be significant.

With these assumptions the terms in the integrand with an overall
time derivative (or two time derivatives) are tractable. They
produce contributions proportional to $\frac{m^2_{\sigma}}{H^2}
\left| \Rcal^{\ixo} \right|^2$ and $(\epsilon-\eta) \left|
\Rcal^{\ixo} \right|^2$. We are then left with three terms in the
integrand, which we rewrite using the same technique as with
$\gamma_\sigma$ in Eq.\ (\ref{eq:gammafinal}). We obtain
\begin{align}
  & 6 H \triangle^{-1} \p_i \left( \dsigmad \p^i \dsigma \right) - 2 \left(
    \dsigmad \right)^2 + m^2_{\sigma} \left( \dsigma \right)^2 \nn\\
  &= \triangle^{-1} \left[ 6 H \p_i \left( \dsigmad \,\p^i \dsigma \right) +
    m^2_\sigma \,\p_i \p^i \left( \dsigma \right)^2 \right] - 2 \left(
    \dsigmad \right)^2 \nn\\
  &= -2 \triangle^{-1} \p_i \left( \delta^{\ixo} \ddot{\sigma} \p^i \dsigma
    \right) - 2 \left( \dsigmad \right)^2~,
\end{align}
where we have used the equation of motion\footnote{We have omitted the spatial
  derivative term $a^{-2} \p_i \p^i \dsigma$ from the equation of motion. It
  may give some contribution immediately after the horizon exit but its total
  effect in the integration over time is negligible.} $\delta^{\ixo}
\ddot{\sigma} + 3 H \dsigmad + m^2_{\sigma} \dsigma = 0$. These
two remaining terms in the integrand we estimate like the terms
outside the integral. They both have two time derivatives and no
overall scale dependence, so we can estimate them roughly as $(
m^2_\sigma / H )^2 \,| \dsigma |^2$. Hence we may write
\begin{align} \label{eq:integralcontribution}
  & \frac{\kappa^2}{\epsilon H} \int \left[ 6 H \triangle^{-1} \p_i \left(
      \dsigmad \p^i \dsigma \right) - 2 \left( \dsigmad \right)^2 +
    m^2_{\sigma} \left( \dsigma \right)^2 \right] dt
  \nn\\
  & \sim \frac{\kappa^2}{\epsilon H} \int \frac{m^4_\sigma}{H^2} \left(
    \dsigma \right)^2 dt \sim \frac{\kappa^2}{\epsilon H}
  \frac{m^4_\sigma}{H^3} \left( \dsigma \right)^2 \int H dt
  \nn\\
  & \sim \triangle N \;\frac{m^4_\sigma}{H^4} \;\left| \Rcal^{\ixo}
  \right|^2~,
\end{align}
where $\triangle N$ denotes the number of e-folds that the scale
in question has spent outside the horizon. The relation $\triangle
N = \int H dt$ follows directly from the definition $N \equiv
\mbox{ln}(a_{\mbox{\small e}}/a)$ where the subscript $e$ denotes
the end of inflation. Since the maximum value of $\triangle N \sim
60$ \cite{liddle00} we can rather safely say that the contribution
from Eq.\ (\ref{eq:integralcontribution}) does not dominate over
$\frac{m^2_{\sigma}}{H^2} \left| \Rcal^{\ixo} \right|^2$.

Hence, combining the contributions of both fields, $\vp$ and
$\sigma$, we may write the final result for the order of magnitude
of non-Gaussianity in hybrid inflation as
\begin{equation}
  \Rcal^{\ixt} = \left( a\,\eta + b\,\epsilon + c_\sigma
  \frac{m^2_{\sigma}}{H^2}\right) \left( \Rcal^{\ixo}\right)^2~,
  \label{eq:finalcurvpert}
\end{equation}
where $a$, $b$ and $c_\sigma$ are constants. Both fields $\vp$ and $\sigma$
contribute to constants $a$ and $b$ but $c_\sigma$ is solely due to the
$\sigma$-field. In principle, the constants could be evaluated using Eqs.\ 
(\ref{eq:elamanvesi}), (\ref{eq:Iphi}), and (\ref{eq:vihonviimeinen}). Simple
estimation suggests they are of the same order. We should however emphasize
that Eq.\ (\ref{eq:finalcurvpert}) should not be taken to suggest that
non-Gaussianity in hybrid inflation is $\chi^2$-distributed; in fact, the
leading terms are non-local, as discussed above.

Thus we may conclude that, barring accidental cancellations in
$\Rcal^{\ixt}_{\sigma}$, the second scalar field $\sigma$ of hybrid inflation
gives the dominant contribution to $\Rcal^{\ixt}$ if the mass of the $\sigma$
field is large enough compared to the slow-roll parameters, i.e.\ if
\begin{equation} \label{eq:condition}
  \frac{m^2_{\sigma}}{H^2} \gtrsim \Ocal (\epsilon, \eta).
\end{equation}
Since usually it is assumed that $\epsilon \ll \eta$ the condition
(\ref{eq:condition}) can be written as
\begin{equation} \label{eq:condition2}
  m^2_{\sigma} \gtrsim \eta H^2~.
\end{equation}

The approach followed in this Section does not apply if $\sigma$
is too massive. When $m_\sigma > 3 H / 2$, the $\sigma$
perturbations acquire a maximum scale dependence $k^3 | \dsigma
|^2 \propto (k/aH)^3$. In addition, the amplitude of perturbations
$|\dsigma|^2$ becomes suppressed by $m_\sigma / H$
\cite{riotto02}. The strong scale dependence is due to the fast
decay of the perturbation outside the horizon, which can easily be
seen from the equation of motion  whose solutions decay
proportionally to $\textrm{exp}(-3 H t / 2)$ when $m_\sigma >
3H/2$. Then immediately after horizon exit $\sigma$ perturbations
would be suppressed as $|\dsigma|^2 \sim (H/m_\sigma)
\,|\dphi|^2$. Thereafter they would decay much faster than the
perturbations of the effectively massless inflaton field.  This
would most likely reduce their contribution to $\Rcal^{\ixt}$
insignificant.

\section{Discussion}

We have presented an analysis of non-Gaussianity in hybrid inflation using a
second order perturbation expansion of the metric and energy-momentum
tensor\footnote{Our formalism is applicable to those multi-field inflation
  models for which the trajectory in the field space is not curved, so that
  one can set the background value $\sigma_0=0$ (the absence of a linear term
  in $\sigma$ is also required); in addition to hybrid inflation, the
  Linde-Mukhanov model \cite{linde96} is another example.}. In hybrid
inflation both the inflaton and the ``waterfall field'' $\sigma$ give rise to
non-Gaussian fluctuations. We showed that the two sources can be separated and
computed the comoving curvature perturbation $\Rcal$ up to second order,
paying particular attention to the contribution of $\sigma$ to $\Rcal$.  We
found that it does not affect the first order perturbation but is present at
second order. Our main observation is that while inflaton-induced
non-Gaussianities always are proportional to the slow-roll parameters, and
therefore usually small, non-Gaussianities induced by $\sigma$ do not have the
slow-roll parameter dependence but rather scale like $m^2_{\sigma}/H^2$.
Hence the waterfall field $\sigma$ is the main source of non-Gaussianity in
hybrid inflation whenever $m^2_{\sigma} \gtrsim \eta H^2$.

In hybrid inflation $H$ is highly adjustable; difficulties start emerging only
when the energy scale during inflation is lowered to $\sim$TeV \cite{lyth99b}.
Therefore no clear-cut conclusion can be drawn for a generic model of hybrid
inflation. Note however that WMAP data for the spectral index $n=0.99 \pm
0.04$ \cite{wmap:parameters} implies that $\eta \simeq 0.01$ through
$n-1\simeq 2\eta$ (assuming $\epsilon \ll \eta$). For illustrative purposes we
could adopt $\eta H^2 \sim ( 10^{9} ~\GeV )^2$ as a typical value for hybrid
inflation. This implies that $\sigma$-induced non-Gaussianities can dominate
already for relatively low $m_{\sigma}$. Moreover, potentially they can be
large. Therefore, the future detection of CMB non-Gaussianities could provide
a powerful tool for testing hybrid inflation models.

Comparison with data is however not straightforward. This is true for any
proper treatment of the second-order perturbations, not just for the case of
hybrid inflation. A fundamental problem is that in the presence of
non-Gaussianities the $N$-point correlators need not be related -- in
principle non-Gaussianity implies an infinite number of degrees of freedom.
Therefore a generic parametrization and observational testing of
non-Gaussianity is difficult (for a discussion, see e.g.\ \cite{komatsu02}).

A measure of non-Gaussianity often used in the literature is the non-linearity
parameter $f_{\mbox{\scriptsize NL}}$ defined by \cite{komatsu02}
$\Phi(\boldsymbol x ) = \Phi_{\mbox{\scriptsize G}}(\boldsymbol x ) +
f_{\mbox{\scriptsize NL}} \left[ \Phi_{\mbox{\scriptsize G}}^2(\boldsymbol x )
  - \left\langle \Phi_{\mbox{\scriptsize G}}^2(\boldsymbol x ) \right\rangle
\right]$, where $\Phi$ is curvature perturbation proportional to $\Rcal$ and
$\Phi_{\mbox{\scriptsize G}} \propto \Rcal^{\ixo}$ is the Gaussian part; the
angle brackets denote statistical ensemble average. By construction this
parametrization is ideal for $\chi^2$ non-Gaussianity. Although the $\sigma$
contribution to $\Rcal$, Eq.\ (\ref{eq:vihonviimeinen}), is of the form first
order perturbation squared, it does not contain an explicit factor
$(\Rcal^{\ixo})^2$ and is therefore not $\chi^2$ distributed. Only the first
term in the inflaton contribution, Eq.\ (\ref{eq:elamanvesi}), is directly
amenable to the parametrization above (and would imply $f_{\mbox{\scriptsize
    NL}} \propto \eta - 3 \epsilon$). However, the rest of the inflaton
contribution, Eq.\ (\ref{eq:Iphi}), together with all of the $\sigma$
contribution, Eq.\ (\ref{eq:vihonviimeinen}), contribute non-locally as they
depend on terms integrated over time or space.

A more appropriate parameter would be the non-Gaussianity kernel $\Kcal$, which
is directly related to the curvature perturbation bispectrum as we will show
below. $\Kcal$ is also used by Acquaviva et al.\ \cite{acquaviva03}. It could
be called a ``momentum dependent'' non-linearity parameter and is defined in
momentum space by \cite{bartolo03a} (see also \cite{acquaviva03})
\begin{align}
  \Phi(\boldsymbol k ) = \Phi_{\mbox{\scriptsize G}}(\boldsymbol k ) +
  \frac{1}{(2 \pi)^3} &\int d^3 k_1 \,d^3k_2 \,\delta^{\ix 3} (\boldsymbol k_1
  + \boldsymbol k_2 - \boldsymbol k) \Kcal (\boldsymbol k_1, \boldsymbol k_2)
  \Phi_{\mbox{\scriptsize G}}(\boldsymbol k_1 ) \Phi_{\mbox{\scriptsize
  G}}(\boldsymbol k_2 ) \nn\\
  & + \mbox{constant}~,
\end{align}
where the constant is such that $\langle \Phi(\boldsymbol k ) \rangle =
0$. With this definition the bispectrum for the primordial curvature
perturbation $\Phi$ reads \cite{bartolo03a,acquaviva03}
\begin{align} \label{bispect}
  \langle \Phi(\boldsymbol k_1) \Phi(\boldsymbol k_2) \Phi(\boldsymbol k_3)
  \rangle = (2 \pi )^3 \delta^{\ix 3} (\boldsymbol k_1 + \boldsymbol k_2 +
  \boldsymbol k_3 ) \,[& 2 \,\Kcal(\boldsymbol k_1, \boldsymbol k_2)
  \,\Pcal_\Phi(\boldsymbol k_1) \,\Pcal_\Phi(\boldsymbol k_2) \nn\\
  &+ \mbox{cyclic} ]~,
\end{align}
where $\Pcal_\Phi(\boldsymbol k)$ is the power spectrum for the primordial
curvature perturbation. The different contributions in Eq.\ (\ref{bispect})
depend on the overall factors in Eq.\ (\ref{eq:finalcurvpert}); hence the
waterfall field $\sigma$ would give the leading contribution to the
non-Gaussianity kernel whenever the condition Eq.\ (\ref{eq:condition2}) is
met.

Although the non-Gaussianity kernel $\Kcal$ is not identical to the
non-linearity parameter $f_{\mbox{\scriptsize NL}}$, we may nevertheless
deduce some rough observational constraints on $f_{\mbox{\scriptsize NL}}$
that can be used to evaluate the significance of the non-Gaussianities
discussed in this paper. For instance, since WMAP \cite{wmap:gaussianity}
established the bounds $-58 < f_{\mbox{\scriptsize NL}} < 134$ with 95\%
confidence, non-Gaussianity given by Eq.\ (\ref{eq:finalcurvpert}) is beyond
the present observational accuracy provided $m_\sigma < H$. Given the cosmic
variance, detector noise, and foreground sources, WMAP is eventually expected
to detect $| f_{\mbox{\scriptsize NL}} |$ with an accuracy of 20, while Planck
has the projected accuracy of 5 \cite{komatsu01}. An ideal experiment has a
sensitivity of 3 for $| f_{\mbox{\scriptsize NL}} |$ \cite{komatsu01}. Hence
there is a possibility that Planck will detect non-Gaussianities if $m_\sigma/
H$ is rather close to unity and the constant $c_\sigma$ in Eq.\ 
(\ref{eq:finalcurvpert}) is larger that unity. Thus a more accurate
determination of the constants $a$, $b$ and $c_\sigma$ in Eq.\ 
(\ref{eq:finalcurvpert}) is clearly desirable.

We should point out that post-inflationary effects provide another source of
uncertainty for a comparison between data and theoretical estimates. Our
formalism, and especially the results, Eqs.\ (\ref{eq:elamanvesi}),
(\ref{eq:vihonviimeinen}) and (\ref{eq:finalcurvpert}), apply only during
inflation. Likewise, the definitions for $\Kcal$ and $f_{\mbox{\scriptsize
    NL}}$ discussed here are for the primordial values, i.e., for the values
before the end of inflation and reheating. We have not considered the effects
of reheating and the later evolution of these perturbations. It is known that
in multi-field models the possibility for generating large non-Gaussianities
after inflation exists; this can happen e.g. in the curvaton scenario (see
e.g.\ \cite{enqvist01, lyth01} for the model and
\cite{lyth02,bartolo03a,bartolo03c} for a discussion on non-Gaussianity). In
such a case one should take the perturbations derived here as initial
conditions for the subsequent evolution, and follow the treatment presented in
\cite{bartolo03b} and applied in \cite{bartolo03a}.

\overskrift{Acknowledgements}

A.V.\ would like to thank Asko Jokinen for useful discussions.  A.V.\ is
supported by the Magnus Ehrnrooth Foundation. K.E.\ is partially supported by
the Academy of Finland grants 75065 and 205800.

\vspace{1cm}
\appendix

\newpage
\storoverskrift{Appendix}

\section{Components of the Einstein tensor} \label{app:einstein}

We take the components of the Einstein tensor, Eq.\ (\ref{eq:einstein-exp}),
from Acquaviva et al.\ \cite{acquaviva03}, where they are presented in a
general gauge. For convenience we cite them here in the generalized
longitudinal gauge for the metric (\ref{eq:metric00})-(\ref{eq:metricij}).

The background components are
\begin{align}
  G^{0 \ix 0}_{\phantom 0 0} &= - \frac{3}{a^2} \left( \frac{a'}{a} \right)^2,
  \\ \nn
  G^{i \ix 0}_{\phantom i j} &= - \frac{1}{a^2} \left[ 2 \frac{a''}{a} - \left(
  \frac{a'}{a} \right)^2 \right] \delta^i_{\phantom i j},
  \\ \nn
  G^{0 \ix 0}_{\phantom 0 i} &= G^{i \ix 0}_{\phantom i 0} = 0.
\end{align}
The first order components are
\begin{align}
  \del G^0_{\phantom 0 0} &= a^{-2} \left[ 6 \left( \frac{a'}{a} \right)^2
    \phi^{\ixo} + 6 \frac{a'}{a} {\psi^{\ixo}}' - 2 \p_i \p^i \psi^{\ixo}
    \right],
  \\ \nn
  \del G^0_{\phantom 0 i} &= a^{-2} \left( -2 \frac{a'}{a} \p_i \phi^{\ixo} -
    2 \p_i {\psi^{\ixo}}' \right),
  \\ \nn
  \del G^i_{\phantom i j} &= a^{-2} \left[ \left( 2 \frac{a'}{a}
    {\phi^{\ixo}}' + 4 \frac{a''}{a} \phi^{\ixo} - 2 \left( \frac{a'}{a}
    \right)^2 \phi^{\ixo} + \p_k \p^k \phi^{\ixo} + 4 \frac{a'}{a}
    {\psi^{\ixo}}' \right. \right. \nonumber\\ \nn
  & \left. \left. \phantom{\left( \frac{a'}{a} \right)^2}
      \quad \quad + \,2 {\psi^{\ixo}}'' - \p_k \p^k \psi^{\ixo} \right)
    \delta^i_{\phantom i j} - \p^i \p_j \phi^{\ixo} + \p^i \p_j \psi^{\ixo}
    \right]~.
\end{align}
The second order components are
\begin{align}
  \ddel G^0_{\phantom 0 0} &= 2 \,a^{-2} \left[ 3 \left( \frac{a'}{a}
    \right)^2 \phi^{\ixt} + 3 \frac{a'}{a} {\psi^{\ixt}}' - \p_i \p^i
    \psi^{\ixt} - 12 \left( \frac{a'}{a} \right)^2 \left( \phi^{\ixo}
    \right)^2 \right. \\ \nn
   & \left. - 12 \frac{a'}{a} \phi^{\ixo} {\psi^{\ixo}}' - 3 \p_i \psi^{\ixo}
    \,\p^i \psi^{\ixo} - 8 \psi^{\ixo} \p_i \p^i \psi^{\ixo} + 12 \frac{a'}{a}
    \psi^{\ixo} {\psi^{\ixo}}' - 3 \left( {\psi^{\ixo}}' \right)^2 \right],
  \\ \nn
  \ddel G^i_{\phantom i 0} &= 2 \,a^{-2} \left( \frac{a'}{a} \p^i \phi^{\ixt}
    + \p^i {\psi^{\ixt}}' + \frac{1}{4} \p_k \p^k {\chi^{i \ixt}}' - 4
    \frac{a'}{a} \phi^{\ixo} \p^i \phi^{\ixo} \right. \nonumber\\ \nn
  & \left. + \,4 \frac{a'}{a} \psi^{\ixo} \p^i \phi^{\ixo} - 2 {\psi^{\ixo}}'
    \p^i \phi^{\ixo} + 4 {\psi^{\ixo}}' \p^i \psi^{\ixo} + 8 \psi^{\ixo} \p^i
    {\psi^{\ixo}}' \right),
  \\ \nn
  \ddel G^0_{\phantom 0 i} &= 2 \,a^{-2} \left( - \frac{a'}{a} \p_i
    \phi^{\ixt} - \p_i {\psi^{\ixt}}' - \frac{1}{4} \p_k \p^k {\chi_i^{\ixt}}'
    + 8 \frac{a'}{a} \phi^{\ixo} \p_i \phi^{\ixo} \right. \nonumber\\ \nn
  & \left. + \,4 \phi^{\ixo} \p_i {\psi^{\ixo}}' + 2 {\psi^{\ixo}}' \p_i
    \phi^{\ixo} - 4 {\psi^{\ixo}}' \p_i \psi^{\ixo} - 4 \psi^{\ixo} \p_i
    {\psi^{\ixo}}' \right),
  \\ \nn
  \ddel G^i_{\phantom i j} &= 2 \,a^{-2} \left\{ \left[ \frac{1}{2} \p_k \p^k
    \phi^{\ixt} + \frac{a'}{a} {\phi^{\ixt}}' + 2 \frac{a''}{a} \phi^{\ixt} -
    \left( \frac{a'}{a} \right)^2 \phi^{\ixt} - \frac{1}{2} \p_k \p^k
    \psi^{\ixt} \right. \right. \nonumber\\ \nn
  &+ \,{\psi^{\ixt}}'' + 2 \frac{a'}{a} {\psi^{\ixt}}' + 4 \left( \frac{a'}{a}
    \right)^2 \left( \phi^{\ixo} \right)^2 - 8 \frac{a''}{a} \left(
    \phi^{\ixo} \right)^2 - 8 \frac{a'}{a} \phi^{\ixo} {\phi^{\ixo}}'
    \nonumber\\ \nn
  &- \,\p_k \phi^{\ixo} \p^k \phi^{\ixo} - 2 \phi^{\ixo} \p_k \p^k \phi^{\ixo}
    - 4 \phi^{\ixo} {\psi^{\ixo}}'' - 2 {\phi^{\ixo}}' {\psi^{\ixo}}' - 8
    \frac{a'}{a} \phi^{\ixo} {\psi^{\ixo}}' \nonumber\\ \nn
  &- \,2 \p_k \psi^{\ixo} \p^k \psi^{\ixo} - 4 \psi^{\ixo} \p_k \p^k
    \psi^{\ixo} + \left( {\psi^{\ixo}}' \right)^2 + 8 \frac{a'}{a} \psi^{\ixo}
    {\psi^{\ixo}}' + 4 \psi^{\ixo} {\psi^{\ixo}}'' \nonumber\\ \nn
  & \left. + \phantom{\frac{1}{2}} 2 \psi^{\ixo} \p_k \p^k \phi^{\ixo} \right]
    \delta^i_{\phantom i j} - \frac{1}{2} \p^i \p_j \phi^{\ixt} + \frac{1}{2}
    \p^i \p_j \psi^{\ixt} - \frac{1}{4} \p_k \p^k \chi^{i \ixt}_{\phantom i j}
    \nonumber\\ \nn
  &+ \,\frac{1}{2} \frac{a'}{a} \left( \p^i {\chi_j^{\ixt}}' + \p_j {\chi^{i
    \ixt}}' + {\chi^{i \ixt}_{\phantom i j}}' \right)  + \frac{1}{4} \left(
    \p^i {\chi_j^{\ixt}}'' + \p_j {\chi^{i \ixt}}'' + {\chi^{i \ixt}_{\phantom
    i j}}'' \right) \nonumber\\ \nn
  & + \,\p^i \phi^{\ixo} \p_j \phi^{\ixo} + 2 \phi^{\ixo} \p^i \p_j
    \phi^{\ixo} - 2 \psi^{\ixo} \p^i \p_j \phi^{\ixo} -  \p^i \psi^{\ixo} \p_j
    \phi^{\ixo} \phantom{\frac{1}{2}} \nonumber\\ \nn
  & \left. - \phantom{\frac{1}{2}} \p^i \phi^{\ixo} \p_j \psi^{\ixo} + 3 \p^i
    \psi^{\ixo} \p_j \psi^{\ixo} + 4 \psi^{\ixo} \p^i \p_j \psi^{\ixo}
    \right\}.
\end{align}

\section{Energy-momentum perturbations of two scalar\\ fields}
\label{app:energymomentumtensor}
    
We consider here the energy-momentum tensor $T^{\mu \nu}$ for two scalar
fields $\vp$ and $\sigma$ minimally coupled to gravity which can be obtained
in a straightforward way by using Eqs.\ (\ref{eq:energymomentumtensor}),
(\ref{eq:phi-expansion}), (\ref{eq:sigma-expansion}) and
(\ref{eq:metric00})-(\ref{eq:metricij}).  The results are ($V_0 \equiv
V(\vp_0,\sigma_0)$)
\begin{eqnarray}
  T^{0 \ix 0}_{\phantom 0 0} &=& a^{-2} \left( -\frac{1}{2} {\vp'_0}^2
    -\frac{1}{2} {\sigma'_0}^2 - a^2 V_0 \right)~,
  \\
  \del T^0_{\phantom 0 0} &=& a^{-2} \left[ -\vp'_0 \dphi' -\sigma'_0
  \dsigma'  + \left( {\vp'_0}^2 + {\sigma'_0}^2 \right) \phi^{\ixo} \right.
  \nn \\
  && \left. \mbox{} - a^2 \left( \frac{\p V}{\p \vp} \dphi + \frac{\p V}{\p
  \sigma} \dsigma \right) \right]~,
  \nn\\
  \ddel T^0_{\phantom 0 0} &=& 2 \,a^{-2} \left\{ -\frac{1}{2} \left( \vp'_0
  \ddphi' + \sigma'_0 \ddsigma' \right) - \frac{1}{2} \left( \left(\dphi'
  \right)^2 + \left(\dsigma' \right)^2 \right) \right.
  \nn\\
  && \mbox{} + 2 \left( \vp'_0 \dphi' + \sigma'_0 \dsigma \right) \phi^{\ixo}
  + \frac{1}{2} \left( {\vp'_0}^2 + {\sigma'_0}^2 \right) \phi^{\ixt}
  \nn\\
  && \mbox{} - 2 \left( {\vp'_0}^2 + {\sigma'_0}^2 \right) \left( \phi^{\ixo}
  \right)^2 - \frac{1}{2} \left( \p_k \dphi \, \p^k \dphi + \p_k \dsigma \,
  \p^k \dsigma \right)
  \nn\\
  && \mbox{} - \frac{a^2}{2} \left[ \frac{\p V}{\p \vp} \ddphi + \frac{\p
  V}{\p \sigma} \ddsigma + 2 \frac{\p^2 V}{\p \vp \,\p \sigma} \dphi \,\dsigma
  \right.
  \nn\\
  && \left. \left. \mbox{} + \frac{\p^2 V}{\p \vp^2} \left( \dphi \right)^2 +
  \frac{\p^2 V}{\p \sigma^2} \left( \dsigma \right)^2 \right] \right\}~, \nn
\end{eqnarray}
\begin{eqnarray}
  T^{0 \ix 0}_{\phantom 0 i} &=& 0~, \\
  \del T^0_{\phantom 0 i} &=& a^{-2} \left( - \vp'_0 \,\p_i \dphi - \sigma'_0
  \,\p_i \dsigma \right)~, \nn \\
  \ddel T^0_{\phantom 0 i} &=& 2 \,a^{-2} \left[ -\frac{1}{2} \left( \vp'_0
  \,\p_i \ddphi + \sigma'_0 \,\p_i \ddsigma \right) - \dphi' \p_i \dphi -
  \dsigma' \p_i \dsigma \right.
  \nn\\
  && \left. \phantom{\frac{1}{2}} + 2 \left( \vp'_0 \,\p_i \dphi + \sigma'_0
  \,\p_i \dsigma \right) \phi^{\ixo} \right]~, \nn
\end{eqnarray}
\begin{eqnarray}
  T^{i \ix 0}_{\phantom 0 0} &=& 0~, \\
  \del T^i_{\phantom 0 0} &=& a^{-2} \left( \vp'_0 \,\p^i \dphi + \sigma'_0
  \,\p^i \dsigma \right)~, \nn\\
  \ddel T^i_{\phantom 0 0} &=& 2 \,a^{-2} \left[ \frac{1}{2} \left( \vp'_0
  \,\p^i \ddphi + \sigma'_0 \,\p^i \ddsigma \right) + \dphi' \p^i \dphi +
  \dsigma' \p^i \dsigma \right.
  \nn\\
  && \left. \phantom{\frac{1}{2}} + 2 \left( \vp'_0 \,\p^i \dphi + \sigma'_0
  \,\p^i \dsigma \right) \psi^{\ixo} \right]~, \nn
\end{eqnarray}
\begin{eqnarray}
  T^{i \ix 0}_{\phantom 0 j} &=& a^{-2} \left[ \frac{1}{2} \left( {\vp'_0}^2
  + {\sigma'_0}^2 \right) - a^2 \, V_0 \right] \delta^i_{\phantom i j}~, \\
  \del T^i_{\phantom 0 j} &=& a^{-2} \left[ \vp'_0 \,\dphi' +\sigma'_0
  \, \dsigma' - \left( {\vp'_0}^2 + {\sigma'_0}^2 \right) \phi^{\ixo}
  \right.
  \nn\\
  && \left. \mbox{} - a^2 \left( \frac{\p V}{\p \vp} \dphi + \frac{\p V}{\p
  \sigma} \dsigma \right) \right] \delta^i_{\phantom i j}~, \nn\\
  \ddel T^i_{\phantom 0 j} &=& 2 \, a^{-2} \left\{ \left[ \frac{1}{2} \left(
  \vp'_0 \ddphi' + \sigma'_0 \ddsigma' \right) + \frac{1}{2} \left( \left(
  \dphi' \right)^2 + \left( \dsigma' \right)^2 \right) \right. \right.
  \nn\\
  && \mbox{} - 2 \left( \vp'_0 \dphi' + \sigma'_0 \dsigma' \right) \phi^{\ixo}
  - \frac{1}{2} \left( {\vp'_0}^2 + {\sigma'_0}^2 \right) \phi^{\ixt} + 2
  \left( {\vp'_0}^2 + {\sigma'_0}^2 \right) \left( \phi^{\ixo} \right)^2
  \nn\\
  && \mbox{} - \frac{1}{2} \left( \p_k \dphi \,\p^k \dphi + \p_k \dsigma
  \,\p^k \dsigma \right) - \frac{a^2}{2} \left( \frac{\p V}{\p \vp} \ddphi +
  \frac{\p V}{\p \sigma} \ddsigma \right.
  \nn\\
  && \left. \left. \mbox{} + \frac{\p^2 V}{\p \vp^2} \left( \dphi \right)^2 +
  \frac{\p^2 V}{\p \sigma^2} \left( \dsigma \right)^2 + 2 \frac{\p^2 V}{\p \vp
  \,\p \sigma} \dphi \,\dsigma \right) \right] \delta^i_{\phantom i j}
  \nn\\
  && \left. \phantom{\frac{1}{2}} + \p^i \dphi \,\p_j \dphi + \p^i \dsigma
  \,\p_j \dsigma \right\}~. \nn
\end{eqnarray}

\end{document}